\documentclass[a4paper,11pt]{article}
\usepackage{amssymb, amsmath, graphicx,epstopdf, amsfonts}
\usepackage[T1]{fontenc}
\usepackage{url}

\begin{document}

\title{Automated Road Traffic Congestion Detection and Alarm Systems: Incorporating V2I communications into ATCSs} 

\author{Vinh Thong Ta\\
University of Central Lancashire (UCLan)\\ 
School of Physical Sciences and Computing (PSC)\\ 
 Preston, UK\\ 
vtta@uclan.ac.uk}


\date{}
\maketitle

\begin{abstract}
In this position paper, we address the problems of automated road congestion detection and alerting systems and their security properties. We review different theoretical adaptive road traffic control approaches, and three widely deployed adaptive traffic control systems (ATCSs), namely, SCATS, SCOOT and InSync.  
We then discuss some related research questions, and the corresponding possible approaches, as well as the adversary model and potential attack scenarios.  Two theoretical concepts of automated road congestion alarm systems (including system architecture, communication protocol, and algorithms) are proposed on top of ATCSs, such as SCATS, SCOOT and InSync, by incorporating secure wireless vehicle-to-infrastructure (V2I) communications. Finally, the security properties of the proposed system have been discussed and analysed using the ProVerif protocol verification tool.      
\end{abstract}

\section{Introduction}
\label{sec:int} 

Nowadays, smart and effective road traffic control and management have become major focus of transportation research due to the increasing amount of traffic congestions and air pollution caused by the rapid growth of vehicles on road. 
There are numerous research and development projects in this field, where different approaches for adaptive traffic light control have been proposed. To deal with heavy road traffic more efficiently at the intersections in metropolises, numerous optimisation algorithms and concepts have been published, some of them have also been broadly deployed. Statistics showed that with adaptive traffic control systems (ATCSs) traffic congestions and, hence, the degree of air pollution can be greatly reduced.  

There are numerous adaptive traffic control systems (ATCSs) and studies  focusing on how road traffic can be optimised and controlled \cite{schutter, tubaishat, koller, rhodes, osorio, shaikh, djahel, leontiadis, curiac, scats, scoot, insync}. However, most of these studies ignore the security issue, but rather concentrate on system design. Further, one of the major weaknesses identified in these works is their limitation in the detection and response mechanisms for traffic congestion caused by unforeseen road incidents (e.g., accident, vehicle breakdowns). To monitor the entire roads  a huge number of roadside sensors or camcorders, cameras needed to be deployed throughout the road, which entails high cost. The reason why traditional traffic control methods are not very effective when dealing with road incident is that there is a rather long path from the incident alert sent by someone who        detects the incident, through the police or authorities and finally arrived at the traffic control authorities. The notification delay can be even larger in case of non-accident incidents such as vehicle breakdown, because people tend to ignore or react slowly in case of non-accident (not critical) situations. 

Looking into the future, the concept of smart vehicles is not just an illusion anymore. 
Numerous research projects, such as EVITA\footnote{E-safety vehicle intrusion protected applications (EVITA), http://www.evita-project.org/} and SeVeCom\footnote{SEcure VEhicle COMmunication,   
http://www.transport-research.info/project/secure-vehicle-communication.} proposed solutions for secure vehicular communications. These projects' main purpose is to secure the wireless vehicle-to-vehicle and vehicle-to-infrastructure communications,  preventing attackers from causing accidents and chaos by eavesdropping and manipulating the communications or stealing secret keys stored inside car modules.  Anonimity of vehicles has been also investigated and solutions proposed. Finally, the concept of self-driving/autonomous vehicles are also heavily  studied as part of R\&D projects, such as the Google self-driving car project\footnote{Google Self-Driving Car Project, https://www.google.com/selfdrivingcar/.}.   

As the vision of smart cities start to become more realistic, incorporating advanced (future) technologies in urban traffic contol systems is an emerging  subject of numerous research projects. 
Following this line, the main focus of this paper is to examine the possibility of applying future vehicular communication technologies to automate the road traffic congestion alerts directly towards the control base stations, which will in turn re-schedule the traffic signals without any human intervention. 

In this position paper, we review the most important works on adaptive traffic control systems (Sections~\ref{sec:literature}-\ref{sec:exist}). Section~\ref{sec:exist} highlights three broadly deployed ATCSs, namely, SCATS~\cite{scats}, SCOOT~\cite{scoot}, and InSync~\cite{insync}. Section~\ref{sec:adversary} discusses the adversary model, in particular, the attackers' ability  and goal in the context of adaptive traffic control systems, along with some potential attack scenarios. We then examine some relevant research questions and directions in this area, as well as the corresponding possible approaches in Section~\ref{sec:questions} and~\ref{sec:approach}, respectively. Finally, in Sections~\ref{sec:S1}-\ref{sec:S2}, we propose two variants of  automated traffic congestion  alarm concept on top of the broadly deployed ATCSs (such as SCATS, SCOOT, and InSync) to increase their efficiency in detecting congestion caused by incidents occur outside their detectors range. Our method assumes smart vehicles, and is based on \textit{vehicle-to-infrastucture communications} rather than vehicle-to-vehicle. The first variant does not rely on the application of precise positioning equipments, while the second assumes that vehicles are equipped with precise positioning devices. Last but not least, we also consider and discuss the security and privacy questions of this system. 

\section{Theoretical Approaches for ATCSs }
\label{sec:literature}
Schutter \cite{schutter} derives an approximate model, which describes lengths of queues as a continuous time function. The author proposed a traffic light switching scheme to describe how the traffic lights can be optimally switched and the system can be controlled. Schutter argues that the proposed switching scheme can efficiently minimise the problem of traffic congestion. Optimisation is used over a fixed number of switch-overs of lights. Furthermore, since the proposed optimisation is performed over number of switch-overs that can be varied dynamically, in case of any emergency incident there is an option to update the average arrival and departure rates of the vehicles.

Tubaishat et. al. \cite{tubaishat}, proposed a decentralised traffic light control using wireless sensor network.  Their system has a three-layer architecture that consists of a traffic flow policy model and a high-level coordination between the Intersection Control Agents (called ICAs). The traffic information is collected and forwarded by wireless sensors deployed on the lanes at  the in and out points of each intersection. These sensors collect different data such as the number of vehicles, their speed, etc., and forward them to the ICAs. The ICAs then determine the optimal flow model of the intersection depending on the data sent by the sensors. 

Koller et. al. \cite{koller} proposed a prototype system based on traffic scene surveillance approach. Namely, a vision based surveillance system was deployed, containing video monitoring systems to monitor the traffic, count the number of vehicles passing through a particular lane, as well as measuring the time of heavy traffic on a particular lane.  

In \cite{rhodes}, Mirchandani et. al. proposed and tested a real time traffic adaptive signal control system (called RHODES). Their proposed base station takes the real time data provided by the street detectors as input for measuring the flow of traffic and then optimally control this flow. The proposed adaptive control architecture is hierarchical, namely, the global traffic control problem is fragmented into sub-problems handled locally, and then are connected to each other in a hierarchical manner to achieve global optimisation. 

In \cite{osorio}, C. Osorio et. al. addressed transportation optimisation problems that accounts for vehicular emission metrics (such as CO emissions) as well as congestion metrics, and proposed a metamodel simulation-based approach. The optimisation framework enables the use of high-resolution microscopic traffic and emissions models for environmental metrics. Their method enables  design of more optimal traffic management schemes. 

In \cite{shaikh}, F. Shaikh et. al. address the problem of traffic density estimation and vehicle classification based on video monitoring systems. They highlight the drawbacks of video based monitoring methods and improve the traffic density calculation by using GPS. Besides, possible solutions for capturing and managing the signals of emergency vehicles are also studied.    

S. Djahel et. al. \cite{djahel} also addressed the problem of transportation optimisation for decreasing the travel time and latency of emergency vehicles. They proposed a traffic management framework in which traffic lights, driving policies, and recommendation for drivers can be dinamically adjusted. 

I. Leontiadis et. al. \cite{leontiadis} evaluated the effectiveness of a decentralized traffic-based navigation system in which instead of distributing traffic information centrally, vehicles report their current information such as location, speed, and travel time to their neighborhood. Their solution assumes vehicle-to-vehicle (V2V) communications and requires that each vehicle be able to act as a traffic sensor, to measure a quantity closely related to the surrounding traffic. Further, vehicles should be able to exchange the sampled information in an ad-hoc manner (i.e.,  vehicular ad-hoc network is assumed). The traffic will then be optimised based on the collected information in a decentralized manner.  A realistic testbed has been setup, and the traffic data recorded in Portland, US, used for evaluation.  Their results show that a decentralized approach can greatly reduce congestion in a realistic scenario.   

Finally, \cite{curiac} D. Curiac and C. Volosencu proposed a hierarchical control architecture for urban traffic system with the goal of optimizing the vehicle and pedestrian traffic flows in big cities. 
The first level of the hierarchy is represented by wireless sensor-actuator networks (WSANs) 
clusters that control the corresponding traffic lights in the related intersections, while The second level is represented by the zonal control systems.  

Besides these theoretical approaches, there are some adaptive traffic control systems (ATCSs) that have been widely deployed, such as SCATS \cite{scats}, SCOOT \cite{scoot} and InSync  \cite{insync}. SCATS is one of the most widely used ATCS, originally proposed and developed by the Roads and Traffic Authority (RTA), Australia. SCOOT is originally proposed and currently widely deployed in the United Kingdom, and inSync is proposed and develop by Rhythm engineering, and broadly applied in the USA.

\section{The Structure and Elements of Current Traffic Control Systems}
\label{sec:strucATCS}
\begin{figure}[htb!]
    \begin{center}
        \includegraphics[width=0.5\textwidth]{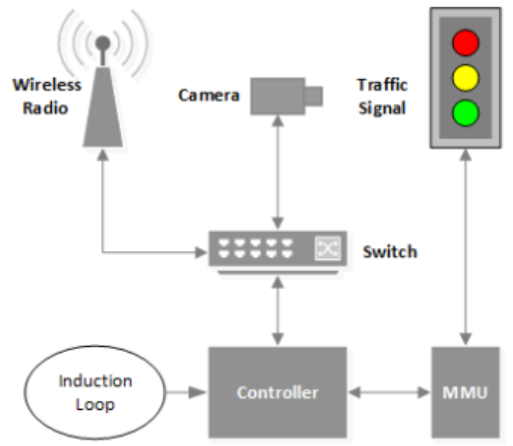}
    \end{center}
    \caption{\textit{A generic traffic light concept.} \cite{forevergreen}}
    \label{fig:TrafficLight}
\end{figure}

The main elements of a typical traffic control system consists of 
sensors, controllers, and networking devices (depicted in Figure~\ref{fig:TrafficLight}).

\textit{Sensors} are mainly used to detect vehicles and measure vehicle related data such as speed. Induction loops are used (by several systems such as SCATS and SCOOT, discussed below) to detect and count vehicles passing certain points of the road. Loop detectors are placed beneath the road, measuring a change in inductance due to the metal body of the vehicle. Video cameras are also broadly installed to detect vehicles and measure their speed. Other methods such as ultrasonic sensor technologies are also proposed\footnote{Institute of Transportation Engineers. Detection trends, 2003.}, however, less used due to their higher cost compared to the induction loop and camcorder solutions.

\textit{Traffic controllers} (Base stations) are connected to the sensors and make calculation and  optimisation on the sensor inputs, as well as  control light states. Controllers are usually placed in a metal cabinet (that provides some degree of physical protection) by the roadside, and  activate the traffic
lights based on relays. There are different operation modes for traffic control, namely 
\begin{enumerate}
\item \textit{pre-timed mode}, when 
lights are controlled only based on preset timings \cite{PhD}. 

\item \textit{semi-actuated} mode, 
where only the secondary (side) streets are activated based on sensors \cite{PhD}.

\item fully-actuated mode, where both types of streets are dealt with based on sensor input \cite{PhD}.
\end{enumerate}

\textit{Networking devices}, such as switch, routers and optical cables ensure communication between the system elements (e.g, controllers-controllers, sensors-controllers, controllers-local base stations). Wired communication (optical, electrical) is usually implemented within a local area, while remote elements communicate with each other based on wireless communication technologies, in either single-hop or multi-hop manner  \cite{forevergreen, ITS}. Radios for traffic control systems usually operate in the ISM band at 900 MHz, 5.8 GHz, or 4.9 GHz \cite{forevergreen, ITS}.

\textit{Malfunction management units (MMUs)} are conflict management units that implement hardware-level safety
mechanisms. They maintain and store a whitelist of safe light states configuration on a
circuit board. Hence, an unsafe configuration (e.g., conflicting green lights) is easily
detected and prevented. MMUs are designed to prevent  critical incidents such as accidents. By implementing hardware-level safety mechanism, MMUs are not susceptible to software security vulnerabilities.     





\section{Real World Traffic Control Systems}
\label{sec:exist} 

In the following, we review the most relevant, not only theoretical but  broadly used and deployed adaptive traffic control systems. We discuss their advantages and limitations, which serves as motivation for our work, namely, investageting how to improve their effectiveness with the potential future (communication) technologies on road. We focus on integrating these traffic control methods with secure  vehicle-to-infrastructre communication concepts. 

In general, adaptive traffic control systems (ATCSs) analyse real time traffic data to optimise the schedule of the traffic light. Their main purpose is to minimise the unused green time and reduce traffic congestion in urban areas. Numerous ATCSs have been developed using different control methods and structure to reduce travel times and congestion \cite{zhao}.

ATCSs are based on efficient optimisation algorithms, hierarchical,  centralized or decentralized architecture. Traditional solutions (e.g., SCATS, SCOOT) optimize the traffic light control based on cycle length, split, offset, and analogue signals \cite{scats1}, while more modern solution (e.g., InSync \cite{insync}) is based on digital signal, as well as advanced video detectors and analysis techniques. 

Despite their effectiveness in case of  normal traffic and scenarios, ATCSs  frequently have issues in dealing with some special aspect like emergency vehicles, pedestrians, accidents \cite{Mladenovic}. 

In the following  subsections, we will review the most relevant and broadly used solutions. We highlight their concepts, effectiveness and shortcomings, as well as the related security issues. 

\subsection{The Sydney Co-ordinated Adaptive Traffic System (SCATS)} 

SCATS is one of the most widely used ATCSs in Australia and the world, developed by the Roads and Traffic Authority (RTA) of New South Wales, Australia in the 1970s \cite{zhao}. 

\begin{figure}[htb!]
    \begin{center}
        \includegraphics[width=0.6\textwidth]{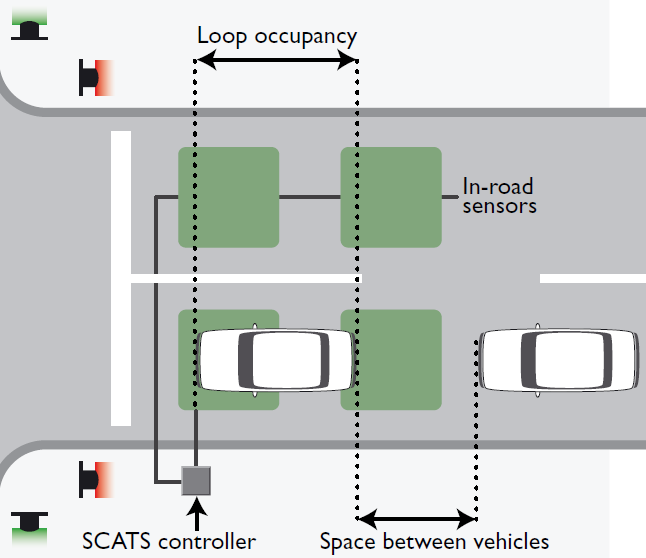}
    \end{center}
    \caption{The SCATS concept based on loop detectors (\cite{scats2}, page 3).}
    \label{fig:scats1}
\end{figure}

\textbf{Concept:} SCATS measures traffic intensity by installing inductive loop detectors beneath the road surface at intersection, right in front of the traffic lights or stop lines (as depicted in Figure~\ref{fig:scats1}).

The induction loops are used to detect the presence of a vehicle to measure the degree of saturation and traffic intensity over a pre-defined time period \cite{Samadi}. Loop detectors (in some cases, video detectors) are installed in every lane at the stop line, which, with the in-road sensors, will measure the distance between the arriving vehicles and the loop occupancy at the stop lines.

The data measured by the induction loops is collected by the local controller (SCATS controller in Figure~\ref{fig:scats1}) which is then transmitted to a regional computer (local base station). After receiving data from each road, the regional computer then  calculate the most ``effective" cycle lengths, splits and offsets for lights in  the area. 

The degree of saturation is the
ratio between demand and discharge flow at traffic light or stop bar. The cycle
length can be set between 20 and 240 seconds, while the
increments of green time are in the range of 4 to 7
seconds \cite{Lowrie}. Split weighting is used for favoring the traffic at the main
intersection, in order to reduce the amount of main road stops. SCATS provides various \textit{manual modes} for control by police or used in emergency situations.

Finally, the calculated schedule is then sent back to the local controllers to adjust the traffic lights accordingly \cite{Dinee}.

\textbf{Effectiveness and deployments:} 
Based on the SCATS brochure \cite{scats2}, on average SCATS has reduced delays by 20\%, reduced stops by 40\%, reduced fuel consumption by 12\% and
emissions by 7\%. 
SCATS is one of the most widely adapted and deployed concept, at about 42000 intersections in over 154 cities and 25 countries including Australia, New Zealand, Hong Kong and Shanghai. 
 
\textbf{Security issues:} In a report on critical infrastructure security \cite{scatssec}, the Australian audit office found that SCATS impelentation provide potential for unauthorised access to sensitive information and systems
that could result in traffic disruptions, or even accidents in the worst case. 

Some of key issues identified by the auditors involving poor SCATS password control; as well as outdated operating systems, and inappropriate anti-virus update management. Auditors were also concerned about the physical security of the roadside cabinets (local controllers), since it were too easy to break into.  

While in the current state, SCATS implements safety interlocks to prevent simultaneous green lights creating a
dangerous situation at an intersection like accidents, a malware infection or unauthorised access can still cause traffic jams, by sending incorrect information/schedule to the local controller.   

\subsection{Split Cycle Offset optimisation Technique (SCOOT)}
Similar to SCATS, SCOOT is a traditional solution based on green-split and offset calculation, originally proposed in the United Kingdom. SCOOT is based on a  centralized architecture and centralized scheduling algorithm. Predefined detection points are installed on the road, which measures the average one-way flow of vehicles. Detectors are installed at the upstream end of the road segments, and a second set of
detector is placed some 50-300 meters before the stop-line. SCOOT maintains the so-called cyclic flow profiles to estimate the number of vehicles that enter the road area each 4-second \cite{Poole}.  


\begin{figure}[htb!]
    \begin{center}
        \includegraphics[width=0.6\textwidth]{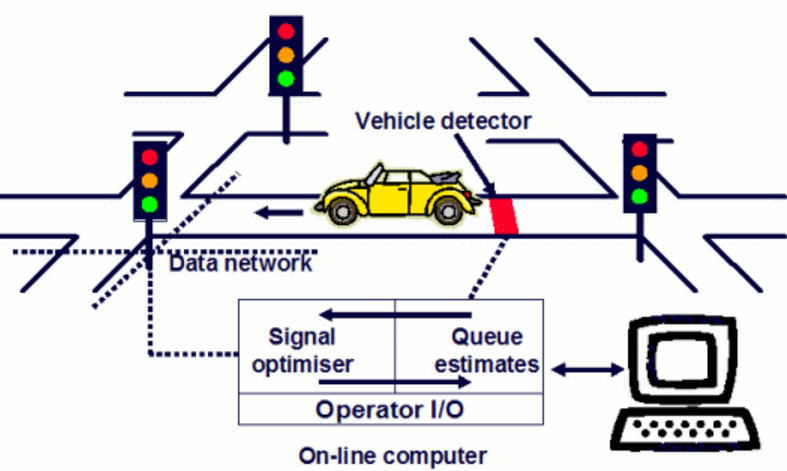}
    \end{center}
    \caption{\textit{The SCOOT concept based on vehicle detectors (mainly induction loop)}. \textit{Vehicles are detected at the start of each approach to every controlled intersection. Profiles are created based on the flows on which the signal optimisation is based.} (Fig. source: http://www.scoot-utc.com/images/HowSCOOTWorks.gif).}
    \label{fig:scoot1}
\end{figure}

SCOOT is based on a queue model, assuming
steady flow during green and optimizes signal for green split, offset and cycle to minimize unused stops and delay \cite{Poole}. These three optimizers are the amount of green for each approach (Split), the time between adjacent signals (Offset) and the time allowed for all approaches to a signalled intersection (Cycle time).

\begin{figure}[htb!]
    \begin{center}
        \includegraphics[width=0.6\textwidth]{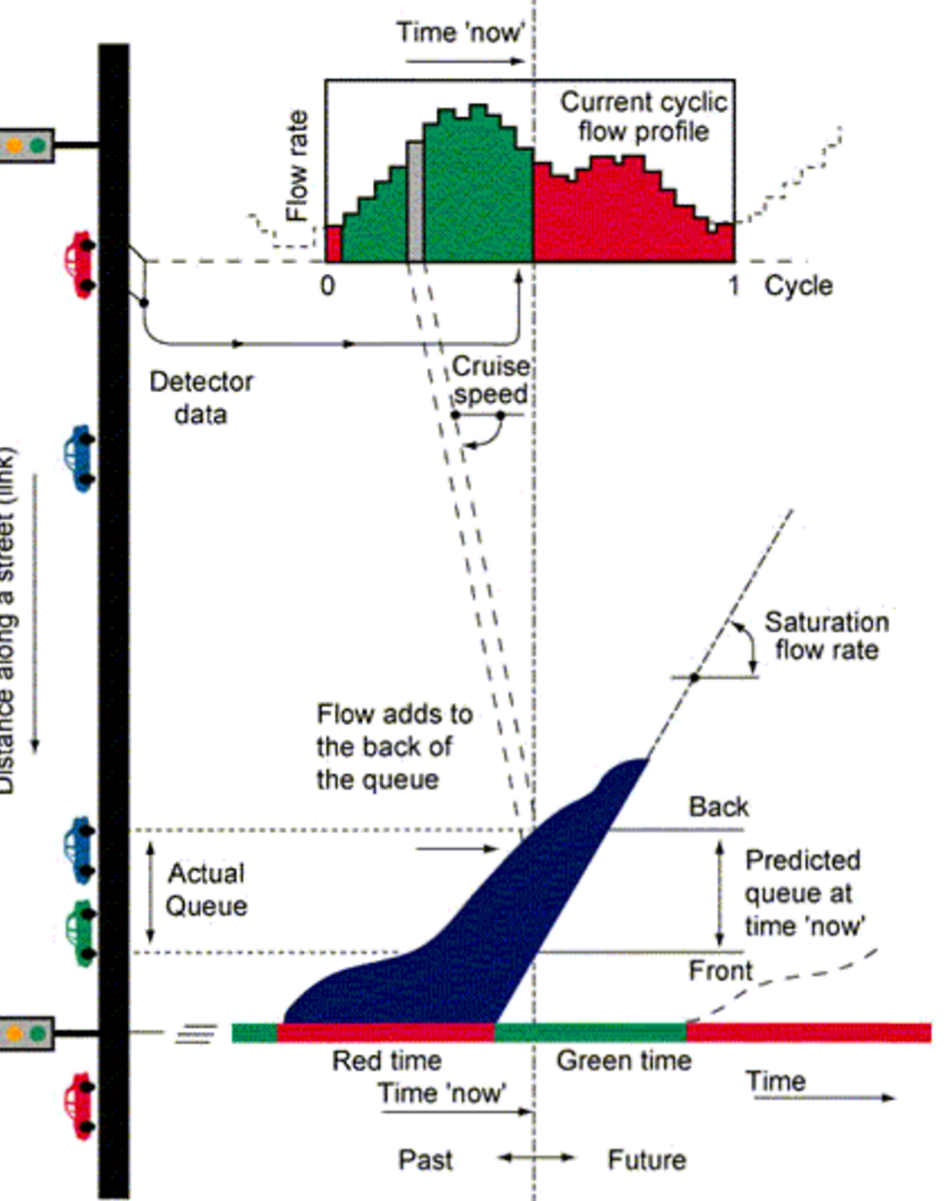}
    \end{center}
    \caption{\textit{When SCOOT receives the information about passing vehicles, it converts the data into its internal units and uses them to construct "Cyclic flow profiles" for each road segment (link). The profile shown on top of the figure is coloured green and red according to the state of the traffic signals when the vehicles will arrive at the stopline at normal speed. Arriving vehicles are added to the back of a (FIFO) queue} (Fig. source: http://www.scoot-utc.com/DetailedHowSCOOTWorks.php).}
    \label{fig:scoot2}
\end{figure} 


Traffic optimisation is carried out in hierarchical levels, namely, Region, Link, Node, and Stage control levels. Region level focuses on cycle length optimisation, Link-level primarily focuses on preventing queue spillback, Node-level performs fine
adjustments of cycle length, offset, and split, while Stage-level provides boundaries for minimum/maximum stage lengths. Optimizers of split, offset, and cycle can be turned on or off, depending on the required level of optimisation. 

Different period of traffic volume is collected, such as 5 or 15-minute, Hourly, Daily and Weekly total volumes, In addition, occupancy levels, queue length, aggregated peak hour flows, and
histograms of flows are also collected, and used in the calculations.

\emph{\textbf{Effectiveness and Deployment:}}
The user interface for SCOOT software is Windows based with specific command language, and supports integration with simulation sofrware such as VISSIM, CORSIM \cite{Poole}. SCOOT has a specialized database called ASTRID (Automatic
SCOOT Traffic Information Database) that is used to store and analyze data such as road intensity, stops, delays, saturation levels, stage length.

A single SCOOT software is able to control up to 300 intersections, and can be extended to 3000 by combining 
multiple computers for operation.
Authors in \cite{Selinger} discussed an estimation about the deployment costs, and made   49,300 USD per intersection for SCOOT.

\emph{\textbf{Security issues:}}
There is not any detailed publication about security analysis of SCOOT, like in case of SCATS, but the literature found that a well-defined and strong access control mechanism is implemented. For instance, user accounts were allowed to be created with up to 16 levels of password access, and only the fully privileged user
can access any level as well as making  modifications to system and user data.
However, like in case of SCATS, the potential of malware infection is present and an interesting issue to examine is how effectively could malwares modify the optimisation routine, causing traffic congestion.  

\subsection{InSync} 
Unlike SCATS and SCOOT, InSync does not rely on loop inductors and splits, offsets optimisation.  InSync was developed by Rhythm Engineering in 2005 \cite{insync}. The InSync control system involves the installation of Internet Protocol (IP) detection cameras at traffic intersections. The cameras are used to detect and quantify the traffic demand situation, in addition to allowing live monitoring of an intersection from an internet browser \cite{insync}. Unlike most traditional control systems, InSync does not use the concept of cycle lengths, splits and offsets, which are all components of the analogue signal control. Instead, InSync system uses the concept of a \textit{finite state machine}. The finite state machine consists of all possible states within the intersection, and specific state can lead to a signal transition/change. 

InSync has three different adaptive stages for local optimisation, namely, phasing, sequencing and green time allocation. In the phasing stage, InSync uses a digital state machine rather than a fixed timing plan. In the sequencing stage, the local traffic engineer can select allowable sequences using CentralSync. Finally, in the green light allocation stage, InSync adjusts green time according to the volume of demand and intersection geometry. 

InSync also relies on a hierarchical structure, namely, it is composed of local and global optimisations. Local optimisation can be overridden at any time by global optimisation \cite{Selinger2}. On the global level, the control system uses the concept of \textit{platoons} (``bunch of vehicles") and focuses on moving these platoons through the selected traffic corridor by manipulating the green time at intersections.  Global optimiser ensures that the intersections interconnect in such way to allow these platoons to travel through with green signals. Local optimiser is responsible for handling the control of the intersections outside of the green time. This concept does not require \textit{the use of a regional computer} like SCATS.

In addtion, the local optimisers are also responsible to handle the volume and delay of individual vehicles instead of the global flow. A greedy algorithm is deployed to limit the time each vehicle must spend at an intersection by applying weightings to each of the vehicles. A corridor with a greater demand of vehicles will have a greater weighting and priority to smaller corridors \cite{Selinger2}. In addition, the local parameters can be used to give higher priorities to special vehicles, such as buses or emergency vehicles.  

\begin{figure}[htb!]
    \begin{center}
        \includegraphics[width=0.7\textwidth]{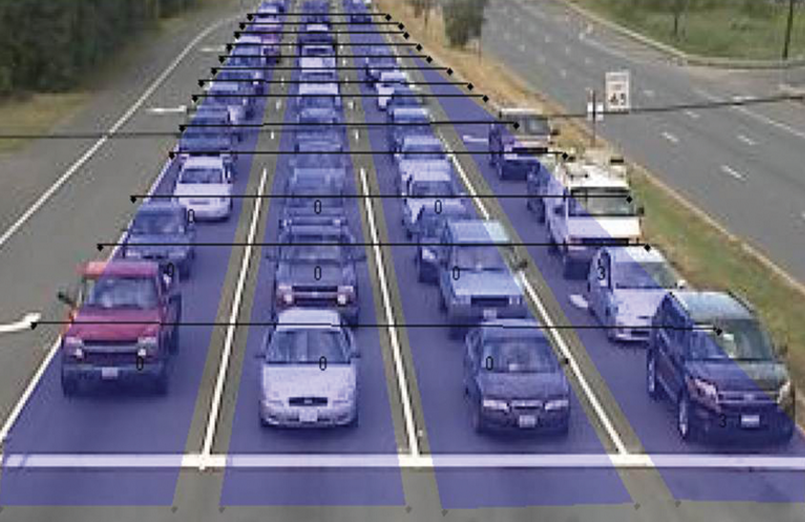}
    \end{center}
    \caption{Vehicle counting concept of  inSync based on cameras \cite{insync}.}
    \label{fig:inSync1}
\end{figure} 
 
The cameras are connected via a secure Ethernet network, and detection zones are drawn along the contours of each lane. The detection zone is futher divided into segments (as depicted in Figure~\ref{fig:inSync1}). Vehicles are counted in front of the traffic lights based on the cameras installed on each lane. By counting how many segments have vehicles in them, InSync constantly monitors and records the number of vehicles in each lane, along with their waiting time.

The system also continuously measures queue and delay at each intersection. Being unconstrained by cycles, the system determines priority so that it can serve approaches from highest priority to lowest priority. If there are a low number of vehicles demanding service, less green time is allocated. For example, by not serving green time to empty approaches and instead distributing time to those approaches with demand. 

\emph{\textbf{Effectiveness and Deployment:}} InSync has mostly been deployed in the US, at  more than 2000 intersections in about 28 states 
\cite{insync}. In 2010, a study was conducted by HDR engineering \cite{Selinger3} in order to assess and compare the performance and costs related to InSync and SCATS. The study found that InSync has the lower cost per intersection, with an average of 28700 USD \cite{zhao}, and better performance in the US. The majority of the intersections within Sydney are fitted more with SCATS than SCOOT or InSync.   

\section{Adversary Model and Possible Attack Scenarios}
\label{sec:adversary}

We examine the attack possibilities based on different perspectives. Note that the following list is not exhaustive, but rather covering our focus.  

The main objectives of the attackers are to make the optimisation working on incorrect data or preventing the optimisation algorithm at the controller running it. As a result, the attackers make the smart control algorithm useless. In particular: 

\begin{itemize}

\item \textit{\textbf{Denial of Service:}} A denial of service (DoS) attack against smart traffic control systems refers to preventing normal light functionality, for instance, setting all
lights to red or green. Denial of service attack may have  critical effect due to the caused traffic congestion. In case the traffic light system is not protected by the hardware based Malfunction Management Unit (MMU) that detects and prevents inconsistent light states, the attackers potentially can bring the traffic lights to an unsafe configuration. However, even if MMU is applied, attackers could make the lights enter a safe but suboptimal state \cite{forevergreen}. Denial of service attacks are easy to detect, and hence, a proper response is given.  

\item \textit{\textbf{Traffic Congestion:}} The attackers' intent is to modify or fake the current traffic data and then get it accepted by the controllers/base stations. As a result, the optimisation algorithm will run on incorrect data leading to congestion. Alternatively, attackers simply prevent  the optimisation from running by infecting the controller/optimisation module with some kind of malware.    
The effect would be that of a poorly
managed road network, again causing traffic congestion. Compared to DoS attack, this type of attack is less detectable, and hence, takes long time to recover.

\item \textit{\textbf{Traffic Diverting:}}  This type of attack is similar to the traffic congestion attack, but here the attackers' goal is to achieve that certain flow is diverted due to the false or incorrect information about traffic congestions on certain roads. This can be critical, for instance, when criminals' objective is to divert certain high-profile vehicles.  
\end{itemize}         

We distinguish   external and internal attackers, as well as local and remote attacks. \textit{Internal attackers} can be further classified into (i) drivers who attempt to modify and hack the vehicles' onboard units such that incorrect messages are sent to regional base stations or roadside units, (ii) technician who have access to the RSUs for doing maintenance tasks (e.g., system update) may be able to track certain vehicles. \textit{External attackers} are neither drivers nor technician/operators who have limited access to either vehicles and roadside units. The main goals of the external attackers are (i) gain unauthorized access (e.g., based on the weakness in physical protection, weak access control, social engineering) to the roadside units, base stations, controller, vehicles onboard computer modules, or (ii) exploting vulnerabilities in the security protocols/primitives used in the wireless communications, either changing the content of messages or replaying old messages.  

On the other hand, \textit{local attacks} are such attacks that either require the attacker to have physical access to the RSUs, controllers or vehicles, or at least in their neighborhood (local area). Local attacks are usually more expensive and easier to detect. On the other hand, \textit{remote attacks} are more convenient for the attackers, and stealth. The attackers can easily erase their trace by using relay proxies, for instance. However, remote attacks usually require infecting the RSUs, controllers, smart vehicles with malware, which will then be controlled by the attacker via command and control servers (C\&C). Even if the system is isolated from the internet, attackers can rely on social engineering to infect the system, e.g., with USB flash drives containing zero-day exploits.                 

\subsection{Published Attacks Against Traffic Lights} 

A study from the University of Michigan (2014) \cite{forevergreen} points out that a large portion of traffic lights in the United States communicated with each other wirelessly over the 900Mhz and 5.8Ghz ISM band \textit{without any encryption}. In order to connect to the 5.8Ghz traffic signals, attackers only need the SSID of the corresponding device. The researchers used the HackRF SDR (Software Defined Radio)  peripheral to sniff and transmit/receive the radio signals from 1 MHz to 6 GHz.

The second security hole the researchers found is that the passwords are left default, which can be found on the traffic light manufacturer's website. In addition to these, to gain access to the 900Mhz networks attackers also need a 16-bit slave ID. The researchers performed brute force search, which was quite easy and fast as no slave ID was greater than 100. After obtaining this, attacker connects to the traffic lights network, with full access to every traffic light connected to the network.

Additional vulnerability can be found in the open debug port in the VxWorks OS which allows the attacker to read-modify-write any memory register.  Another attack is called the remote keypad you can freeze the current intersection state, modify the signal timing, or change the state of any light. However,  the hardware based Malfunction Management Unit (MMU) will still detect any illegal states (conflicting green or yellow lights), and take over with the familiar 4-way red flashing. Since a technician will have to come out and manually reset the traffic signal to recover from an illegal state, the attacker could turn every intersection on the network into a 4-way stop, causing traffic jam.


\section{Research Questions}
\label{sec:questions}

\subsection{Limitation of the Discussed Adaptive Traffic Control Methods}

Current methods are efficient in handling normal traffic flow, while they are less effective in case of abnormal situation such as unforeseen vehicles breakdown, accidents, or other emergency scenarios, which need rapid response. Since the discussed methods are already deployed in several countries with high cost, it is not reasonable to completely replace them.  
Instead of proposing completely different traffic optimisation algorithms, we extend these methods with secure  vehicle-to-infrastructure (V2I) communications to improve their efficiency in emergency and unforeseen scenarios. 

\begin{figure}[htb!]
    \begin{center}
        \includegraphics[width=1\textwidth]{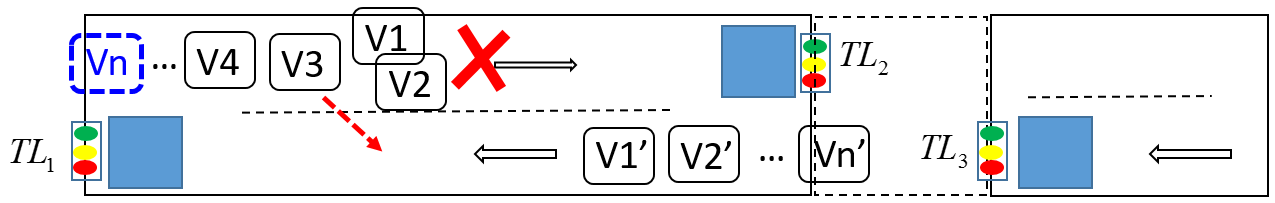}
    \end{center}
    \caption{An unforeseen incident causing traffic jam in a road with SCATS deployed. The (blue) boxes at the traffic lights represesing the loop detectors, while \textit{Vi} denotes a vehicle.}
    \label{fig:accident1}
\end{figure}

Figure~\ref{fig:accident1} shows an example scenario, where SCATS has been deployed on a long road with two lanes (reverse directions). The loop detectors are denoted by the blue boxes in front of the traffic lights. 
Assume that there is an accident between vehicles \textit{V1} and \textit{V2} on the left lane, and hence, all the vehicles behind them \textit{V3},\dots, \textit{Vn} got stuck due to the dense opposite traffic on the right lane (a vehicle breakdown can also result a similar congestion). The situation is more critical when \textit{Vn} is an emergency vehicle on action. Since the road is long, the reaction time  to get information about the accident, and hence, the traffic light re-scheduling can be slow in case of SCATS.  The situation is similar in case of  SCOOT or even InSync where the cameras installed at the traffic lights will not be able to detect the accident occured outside its scanning range. 

\subsection{Related Research Problems}
\label{sec:directions}
In this paper, we focus mainly on the research questions related to effective road congestion detection and alarm  technologies. It has been shown that, currently, most vehicle detection systems apply video detection or induction loops, and most of the countries apply one of the discussed adaptive traffic control methods. Hence, the most cost effective solution would be not to replace entirely the recent deployments, but proposing some extension on top of these solutions, for instance, with modern vehicular communication technologies, or incorporating the concept of social networks/community mapping into the vehicular network context to improve the road congestion detection and alarm  mechanisms.  In the following, we outline some interesting related research problems, and will focus on the \textit{Problem 3} in the rest of the paper. 

\textit{\textbf{Problem 1:}} By involving vehicle-to-infrastructure (V2I)  communications, an additional issue we have to face is the anonymity of the vehicles, namely, to make vehicle tracking more complicated or practically infeasible. Another problem we have to consider is the potential  false alarm coming from hacked vehicles, or vehicles infected with malwares.            

\textit{\textbf{Problem 2:}} 
As we discussed previously, the (regional and global) controllers as well as the sensors are computer based, some of them with windows operating system installed. This entails potential of virus and malware infections. Even in case antivirus can be installed in these controllers, the possibility of zero-day vulnerability and exploitation (e.g., through USB flash drive) is still possible. Although the MMU is designed to prevent the controllers from arbitrary change the state of traffic lights, and since it is not sofware-based, malware still could change the optimisation algorithm at the controller or prevent it from running and hence, causing traffic congestion. 

Hence, an interesting question would be the infection possibility of the roadside sensors and cameras. Further, if the traffic control systems are susceptible to infection with  malwares, especially zero-day malwares, which cannot be detected efficiently by antivirus, then how to detect traffic data anomalies? Both sensors and cameras have specific hardware/software architecture, and hence, writing specific malware for them can be challenging but not impossible. The authors in \cite{sensormal}, for instance, presented some interesting code injection attacks against embedded devices with Harvard architecture. On the other hand, roadside sensors can be physically protected which complicates the task of the attackers. Therefore, the attackers have to rely on remote exploitation.  

An another approach is not to infect the sensors with malwares but exploiting the vulnerabilities found in the communication protocol between the sensors and controllers, either by replaying the old messages of the sensors to the controller or modifying the message elements. This can be prevented by using appropriate cryptographic primitives.       

\textit{\textbf{Problem 3:}} What kind of devices, architecture and communication protocols are required to enable automated road congestion detection and alarm without any human intervention? By without human intervention, we mean the detection and alarm mechanism based entirely on the vehicles and road side infrastructure. We can assume smart vehicles and roadside sensors, and the communication between them. We also distinguish the cases when vehicles are or are not equipped with any GPS  or precise positioning devices, as well as camcorders or lane detectors, respectively. For instance, self-driving vehicles are usually equipped with cameras for capturing pictures about the environment \cite{googlecar}, which can be used to detect road incidents.          

\textit{\textbf{Problem 4:}} How social networks concept could be applied to the smart traffic control? Waze\footnote{Waze, https://www.waze.com/en-GB/about}, a GPS and community based navigation application has been broadly used recently all around the world. Users can share traffic status on certain road segments with each other, such as indicating traffic jams, police presence, constructions, etc. Users can also take  photos about the road incidents and post them. Based on this, drivers can take appropriate action, for instance, slow down to avoid penalty, or just choose an  another road to avoid traffic jam. To mitigate the effect of false information, drivers can also remove the alerts set by other drivers.  It would be an interesting area to investigate how this community concept can be incorporated into automated incident (congestion) alert towards the traffic control base stations, regional controllers, as well as how the optimisation could take into account the alerts sent by drivers. On the other hand, without a proper security solution, criminals could easily mislead the base stations by, e.g., hiring people to send incorrect incident alerts (this activity also known as crowdturfing \cite{turfing}). This is confirmed by the fact that attack possibilities against Waze \cite{wazeattack, wazeattack2} have been published.    
    
\section{Possible Approaches}
\label{sec:approach}
In the following, we discuss some possible solutions for the problems 1-4 highlighted in Section~\ref{sec:directions}. Note that we do not attempt to be complete, but rather, collect the most representative approaches, based on the literature.

\textit{\textbf{Problem 1}}:  In secure vehicular communications concepts, messages sent by the vehicles and the roadside units (RSUs) are usually digitally signed to provide authenticity. The signing keys of the vehicles and the corresponding certificates would make the vehicles become traceable. Hence, to achieve a certain degree of anonymity and privacy for vehicles, several solutions have been proposed in the literature, such as group signature \cite{group} and short-term pseudonyms \cite{sevecom2}. The idea behind group signature is to allow a vehicle to sign messages anonymously on behalf of a group of vehicles, while pseudonyms are short-term vehicle credentials including short-term keypairs used for signature generations/verifications. Vehicles regularly change their pseudonyms making tracking more difficult.     


\textit{\textbf{Problem 2}}: As we discussed, if the elements of the smart road traffic control system (such as sensors, camcorders, RSUs) can be infected with malwares, including zero-day malwares, then they can send incorrect traffic information to the controllers and base stations. In case the base stations rely entirely on the information sent by the sensors, RSUs, and camcorders, it is very difficult to perform automated anomaly detection. A possible solution could be using machine learning methods on the past traffic patterns on certain road segments. The main challenge we have to face in this case is to keep the level of false positive and negative as low as possible. Further, adversarial machine learning methods \cite{adversarial} may be required to cope with adaptive attackers, who attempt to either mimic normal patterns or pollute the training dataset.     

\textit{\textbf{Problem 3}}: Since location information about vehicles will indicate the location of the incident, to enable automated road incident  alerting system, extremely precise information about recent vehicle locations is required.  Therefore, in the following, we distinguish two possible settings: (\textit{S1}) the first setting assumes vehicles without precise positioning equipment installed (including satellite navigation equipments), (\textit{S2}) the second settings assume vehicles with precise positioning equipments. These equipments are able to send information about the road segments and lane on which the vehicles are currently travelling/standing. 

\textit{\textbf{Problem 4}}:  Although our focus in this paper is mainly on automated alert mechanism, problem 4 discusses a very interesting and complex problem since it involves human aspects, involving internal attackers who are basically parts of the system (drivers). This provides space for some interesting areas and approaches such as game theory and machine learning. A possible direction could be incorporating the social concept of community mapping services (such as Waze) to the adaptive traffic control system. Namely, the regional controllers and local base stations  receive road incident information sent by drivers and then incorporate the received data into the optimisation process. From security perspectives, detecting misbehaving drivers or anomalous traffic data could be our major focus. Solution applied in vehicular ad-hoc networks to detect misbehaving nodes, such as majority voting/decision \cite{majority} or machine learning methods seems to be promising approaches. 

In the upcoming sections, we focus mainly on the \textit{Problem 3}, and investigate possible concepts for fully automated traffic congestions alarm systems, by incorporating vehicle-to-infrastucture communication technologies.

\section{A Secure Traffic Congestion  Alert Concept (Type \textit{S1})}
\label{sec:S1}

We propose a solution on top of the existing ATCSs, by installing roadside units (RSU) on the road segments and  the onboard modules in the vehicles enabling secure communications between the vehicles and the RSUs, as well as vehicles and the regional control stations (vehicle-to-infrastructure communications). Let us consider again the example scenario in Figure~\ref{fig:accident1}. In our method, the vehicles behind \textit{V1} and \textit{V2} will simultanously send their information about the obstacle to the local base station, enabling a faster traffic re-schedule. The sent information can include the presence of emergency vehicles as well. The response could be, for instance, extending the green period of either $TL_1$ or $TL_2$ and the red period of $TL_3$. In other words, we base on the crowdsourcing activities of vehicles to enable automated road congestion alert and response. This section introduces a possible approach of type \textit{S1}, where we assume vehicles \textit{without} any precise positioning equipment installed (including satellite navigation devices).  The reason of considering this option is because the precise (satellite) positioning equipments are quite expensive (mainly applied in ther   army) and GPS is still error-prone  when we need to identify the lane and position of a vehicle at a particular time.   

\subsection{Smart Vehicles and Secure Wireless Communications}






Secure communications between vehicles and roadside units (RSUs) are based on digital signature. 
We assume a Public Key Infrastructure for vehicular systems (VPKI), where communications between vehicle with RSUs are digitally signed, and Certificate Authorities (CA) is responsible for issuing and revoking certificates for the public keys.     
At the beginning, each vehicle is registered with a CA, and given (i) a \textit{unique long-term identity}, (ii) \textit{a pair of private and public cryptographic keys}, and (iii) \textit{a
long-term certificate} that contains parameters about the vehicle and the information about the issuing CA.  

Vehicles and RSUs, mRSUs (see below) are equipped with a Hardware Security Module (HSM) and an on-board unit (OBU). The HSM stores and physically \textit{protect private keys for digital signatures} and \textit{provide a secure time source for timestamps}. The installed HSM is ideally tamper-resistant and tampter-proof, meaning that any physical intrusion to the device will be detected with a proper response (e.g., deleting the private keys in the storage). This helps preventing the attacker from stealing private keys to sign messages in the name of a vehicle or RSU. HSM is also responsible for computing digital signatures, while OBU is used for verifying signatures coming from the RSUs, and hence, they are not tamper-proof/resistant like the HSMs.     

To prevent RSUs from being able to track the vehicles, vehicles use \textit{short-term key pairs} in the communications instead of the long-term key pairs. This kind of solution is referred to as ``pseudonymous authentication" \cite{sevecom1, sevecom2}. The CAs are given ability to link long-term identities with several corresponding short-term credentials to provide accountability. The short-term public keys (pseudonyms) do not reveal the vehicles identity, and each vehicle will switch to an another (not previously used) short-term key pair at each intersections. Intersections prevents the observer/attacker from tracking a certain vehicle due to the crowd, and the fact that messages signed under different short-term private keys cannot be linked. 

Each vehicle \textit{VE} has a unique
long-term identity \textit{ID}$_{VE}$, a long-term key pair, and long-term certificate, which is an 
agreement between car manufacturers and the corresponding CA. 
Besides, at regular time interval for each \textit{VE}, a set of $\{(SK^{Pseu1}_{VE}$; $PK^{Pseu2}_{VE}$); \dots; ($SK^{Pseuj}_{VE}$; $PK^{Pseuj}_{VE}$)\} will be generated by its HSM. The long-term credential is used by \textit{VE} to authenticate itself to the CA when requesting a new set of short-term keys pairs (at the intersections). 

\begin{figure}[htb!]
    \begin{center}
        \includegraphics[width=1\textwidth]{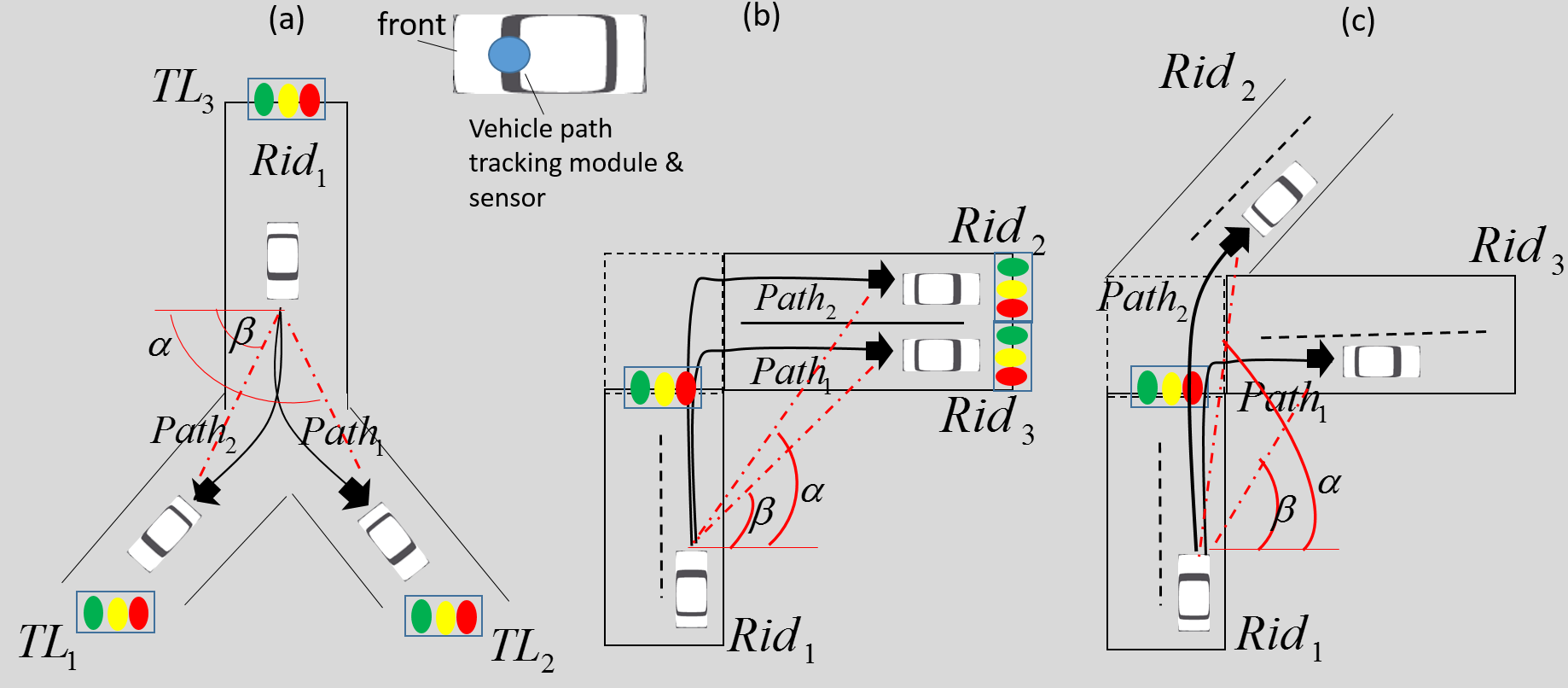}
    \end{center}
    \caption{\textit{Vehicle path recorded by its onboard module/sensor and the expected lane and road on which the vehicle currently is. The path is calculated (drawn) based on the wheel position (tilt angle) and the speed of the vehicle, without the need of precise GPS/positioning devices}.}
    \label{fig:VEpath}
\end{figure} 

\textbf{\textit{Vehicle Path and Lane Estimation}.} We assume that vehicles onboard units (OBU) are also equipped with a module/sensor for tracking and recording the recent portion of the path travelled. The recent portion is recorded from the point a vehicle arrived to a new road segment (reset \textit{dist} performed) to the same point in the following road segment. The previous path will be deleted once the current path started to be recorded. In Figure          ~\ref{fig:VEpath}/(a), \textit{Path}$_1$ and \textit{Path}$_2$ prepresent two paths when the vehicle follows the left and right branch of road segment. To distinguish the two paths (determining if the vehicle is on the left or the right lane), the angle between the horizontal line and the line linking the start and end points of the path can be used. The angle $\beta$ should be an acute angle, while $\alpha$ should be an obtuse angle. Similarly, Figure~\ref{fig:VEpath}/(b) shows a more interesting scenario when roads \textit{Rid}$_{2}$ and \textit{Rid}$_{3}$ are very close to each other, however, in this case, the ranges of the angles $\alpha$ and $\beta$  can be used to determine the lane on which the vehicle currently is. The situation is similar in case of Figure~\ref{fig:VEpath}/(c). 
Figure~\ref{fig:VEpath}/(a) also shows that a single road segment can be branched, as we only consider the segment from one traffic light to another traffic light, stops and intersections without traffic light do not divide the segments.  

\textit{\textbf{Alternative solutions}}: Besides the proposed vehicle path and lane estimation solution above, there are 
several alternative solutions. For instance: 
\begin{enumerate}
\item \textit{Installing camcorders} at the start of each road segment which scan the license plate number of the vehicles arriving at the road. One camcorder is installed on each lane of the road.  The RSUs then use the scanned plate number to  communicate with the vehicles. The vehicles only deal with the message includes their plate numbers. To track the lane change of the vehicles camcorders, should be installed  throughout the road segments at a regular distance. 

This method, however, has several  disadvantages, such as the cost of installing high number of camcorders. Further, the efficiency of license plate number scan depends on the quality of the light and the motion of vehicles, which is error-prone. Third, this solution allows the (colluding) RSUs to track the vehicles based on their plate numbers. Finally, attackers can slightly manipulate the license plate number to mislead the scanning result. 

\item An another possible solution is based on \textit{road surface marking/painting}.  Each lane contains certain road surface marking with the information  about the particular road segment and lane. Vehicles are equipped with the camcorders similar to the reverse cameras   which is used to scan the surface marks. Afterwards, use the scanned information to communicate with the RSUs, signalling their status. Furthermore, lane detection and lane departure warning technologies  can be applied to trace the lane changes of a vehicle.   

The main disadvantage of this solution is that it depends strongly on the quality of the painting on the road and the light condition. In addition, 
it cannot be applied, for instance, in heavy snowing weather. Finally, since the marks are accessible to anyone, attackers can manipulate the surface painting marks, causing errors in the scanning process. 
 
\item A similar solution is based on \textit{road sign boards}, where the boards contain signs encoding the information about the road segments and the lanes. Vehicles are equipped with camcorders that scan the codes on the taffic boards, and then using it in the follow up communications. 
  
This solution is slightly better than the surface marks as it is more effective in different weather conditions, but also susceptible to attacks based on modifying the boards leading to failed scanning.       
Although the concept of scanning the road environment is already applied in self-driving cars (e.g., Google car), recently it is not well accepted by the laws in numerous countries due to the privacy problems.
   
\end{enumerate}

\subsection{Road Segments and Roadside Units}

\begin{figure}[htb!]
    \begin{center}
        \includegraphics[width=1\textwidth]{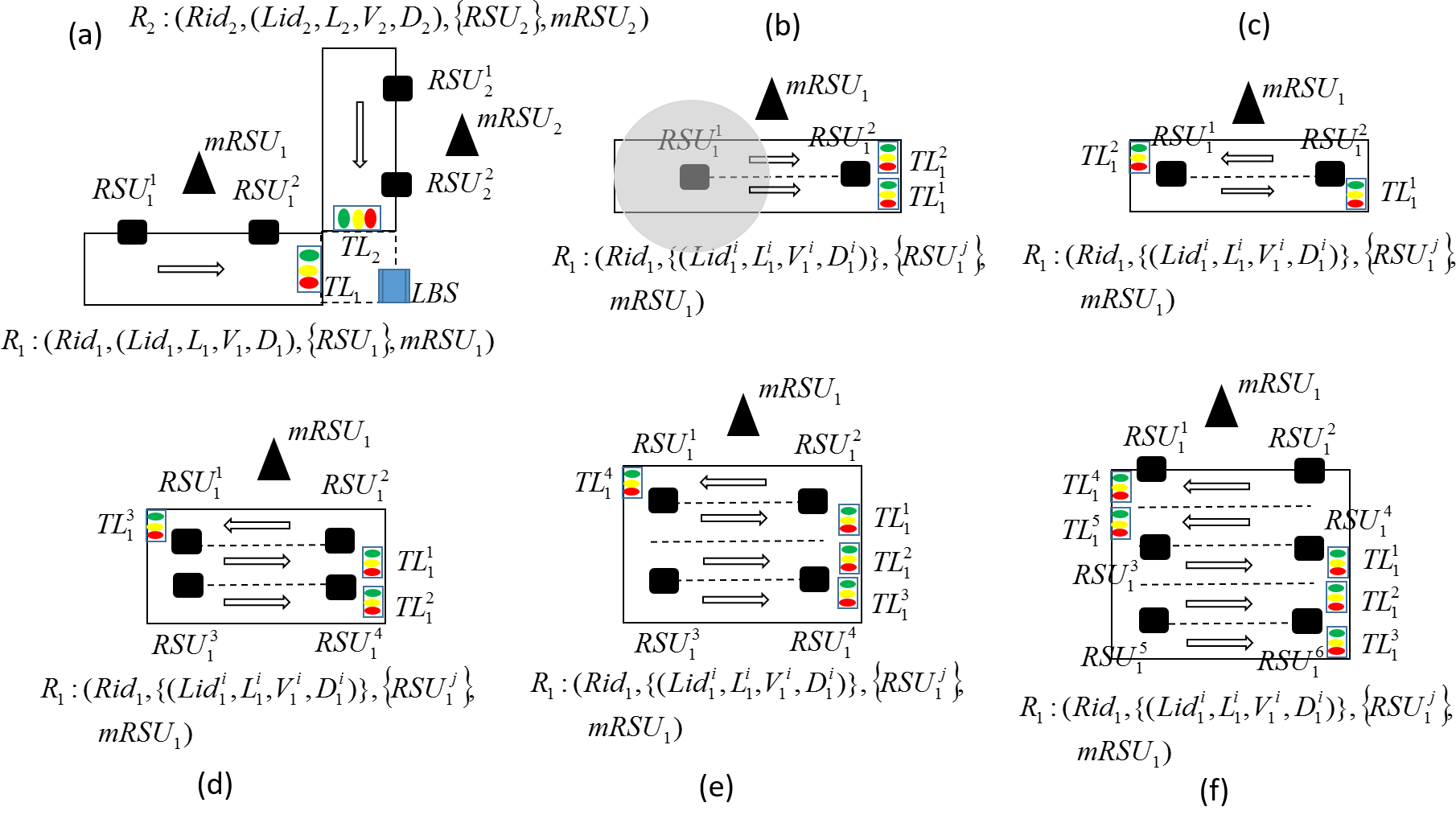}
    \end{center}
    \caption{\textit{An overview of road segments with the RSUs, mRSUs, and LBS.}}
    \label{fig:concept2}
\end{figure}

The basic concept is depicted in the Figure~\ref{fig:concept2}. In our model, we consider (road) segments as the part from a junction with traffic lights to an another junction with traffic lights, and denote them by $R_i$, $i$ $\in$ \{$1$, \dots, $n$\}. We note that if there is a junction without traffic light, we extend the segments until the nearest junction with traffic light. For instance, the Figure~\ref{fig:concept2}/(b)-(f) show a segment $R_1$ with the ID \textit{Rid}$_1$ and  two lanes. Each lane is defined by the 4-tuple (\textit{Lid}$^i_1$, $L^i_1$, $V^i_1$, $D^i_1$), where  \textit{Lid}$^i_1$ is the ID of the lane, $L^i_1$ is the length of the lane, $V^i_1$ represents the \textit{average} speed limit (there can be different speed limit in the same lane, we consider the average of them), $D^i_1$ is the direction of the lane that can be right or left \{$R$, $L$\}, respectively, and finally, a set of ``secondary" roadside units \{\textit{RSU}$^{j}_{1}$\}, and a main roadside unit \textit{mRSU}$_{1}$. The main RSUs are more powerful than the secondary RSUs, since they have to deal with a huge number of messages coming from vehicles.   

Figure~\ref{fig:concept2}/(b)-(c) show a road with two lanes of same and reverse directions, respectively. We assume that each RSU's communication range has a radius of two lanes width (as depicted in Figure~\ref{fig:concept2}/(b)), hence, for the case of (b) and (c)  only two RSUs are required.  Figure~\ref{fig:concept2}/(d)-(f) show examples of road with more lanes, where four RSUs are required to cover the entire width of the road. The traffic lights at the end of the lanes are denoted by \textit{TL}$^{i}_{1}$ for different $i$. The road segments with additional lanes are similarly defined.  Figure~\ref{fig:concept2}/(a) highlights a very simple intersection with two segments $R_1$ and $R_2$ with one lane, and traffic lights \textit{TL}$_{1}$, \textit{TL}$_{2}$. 

The RSUs are placed close to the traffic lights each at the two ends of the road segment, as depicted in the figures. 
Each RSU has a short-range radio antenna that enables reception and transmission of data within the radius equal to the width of two lanes. For this purpose, a similar antenna can also be deployed in the RSUs as the Remote Keyless System (RKS) used in modern vehicles that usually has the range 5-20 meters. Alternatively, ISM band wireless communication can be applied. The reason of using short-range antenna is because we would like to minimise the number of messages that a single RSU has to deal with (due to a huge number of broadcast  messages sent by the vehicles). A vehicle only starts to communicate with a RSU at the beginning/end of a corresponding road segment when it is about to enter/exit that segment (i.e., when it gets close to the RSU). Furthermore, the placement of the RSUs should minimise (or totally exclude) the intersection of the communication range of RSUs belonging to different \textit{Rid}s (road segments) reducing the inconsistencies in the messages received by a vehicle. 

\begin{figure}[htb!]
    \begin{center}
        \includegraphics[width=0.8\textwidth]{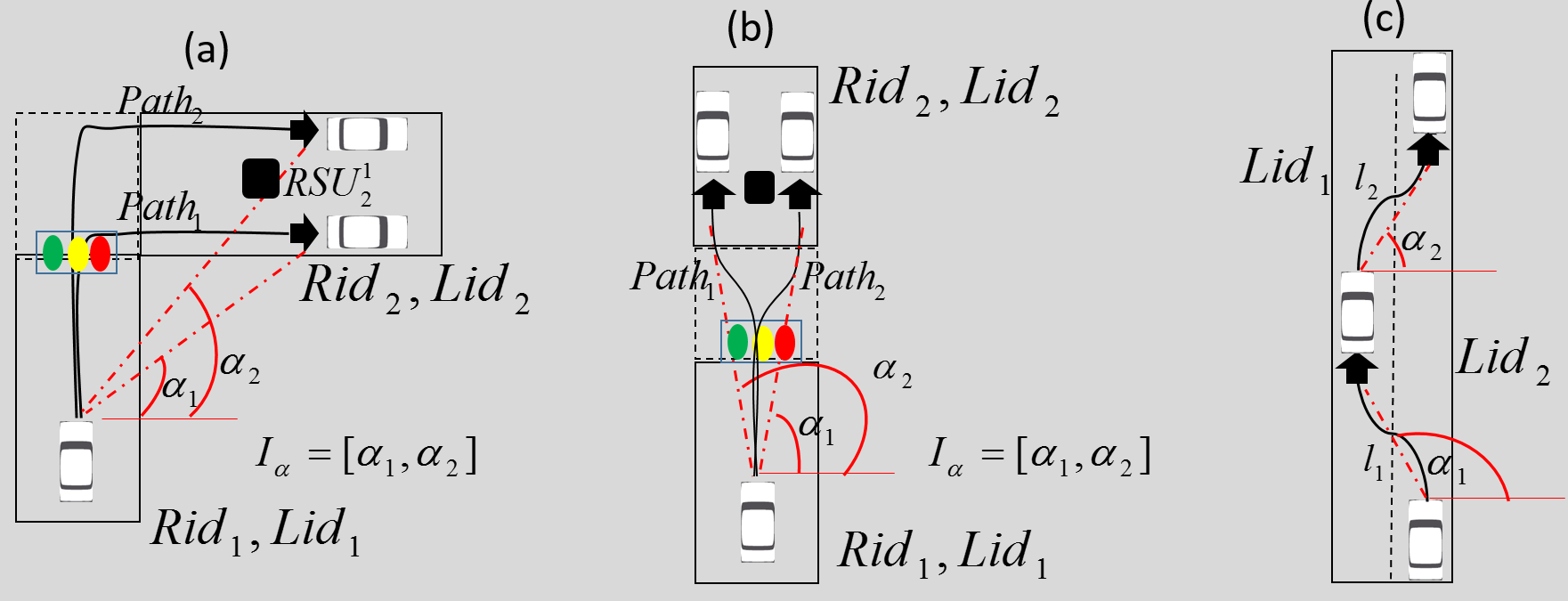}
    \end{center}
    \caption{\textit{The figures show examples on the identification of the lane to which a vehicle has arrived, based on a certain movement angle interval [$\alpha_1$, $\alpha_2$]. Figure (a) and (b) show example for different road segments, while Figure (c) shows lane changes in the same road segment.}}
    \label{fig:angleint}
\end{figure}

Each RSU is pre-installed with information related to a particular road segment. Each RSU deployed in \textit{Rid}$_i$ stores information about the surrounding roads, namely, a set of 4-tuples (\textit{Rid}$_j$, \textit{Lid}$_j$, \textit{I}$_{\alpha}$, \textit{Lid}$_i$), where \textit{I}$_{\alpha}$ is a range of angles $\alpha$ that a vehicle arriving from road segment \textit{Rid}$_j$ and lane  \textit{Lid}$_j$ to the lane \textit{Lid}$_i$ of \textit{Rid}$_i$ is expected to make. For instance, Figures~\ref{fig:angleint}/(a)-(b) represent road segments with only one lane. In Figure~\ref{fig:angleint}/(a), \textit{RSU}$^1_2$ on road segment \textit{Rid}$_2$ stores (\textit{Rid}$_1$,\textit{Lid}$_1$, [$\alpha_1$, $\alpha_2$], \textit{Lid}$_2$). Each RSU also stores information about the road segment ID and lane ID(s), and the direction of the lane for arriving vehicles. For instance, in Figure~\ref{fig:angleint}/(a), \textit{RSU}$^1_2$ stores \textit{Rid}$_2$, \textit{Lid}$_2$. A similar example is depicted in Figure~\ref{fig:angleint}/(b) with the straight road segment \textit{Rid}$_2$. 
Finally, Figure~\ref{fig:angleint}/(c) describes the lane changes of the same vehicle on a road segment. Depending on the width of the lanes, the scale of vehicle movements $l_1$ and $l_2$ (i.e., the length of the red dotted lines), and the angles $\alpha_1$ and $\alpha_2$, the lane on which the vehicle currently travels, can be estimated. For instance, changing from the right lane to the left  is characterized with the obstuse angle ($\alpha_1$), and the opposite with the acute angle.   
The same concept can be applied to the roads with several lanes, where the RSUs store such information of each lane in the road segments.  

Finally, we assume a local base station LBS installed at each intersection (see Figure~\ref{fig:concept2}/(a)). LBSs are responsible for optimizing and re-scheduling the traffic lights accordingly at the dedicated intersection. LBSs can forward information about the traffic flow and the incident alerts to the other LBSs in the neighborhood, extending the optimisation. Note that instead of installing a LBS at each intersection we can also apply regional LSBs involving several intersections. However, in this case LBS should be more powerful to make calculation and optimisation for the whole region, as well as mRSUs also need more powerful antenna in order to transmit messages to a LBS far away. The communications between vehicles and mRSUs are also based on longer range radio communication.      

\subsection{Automated Congestion Alarm  Algorithm Without GPS/Precise Positioning Devices Installed} 
\begin{figure}[htb!]
    \begin{center}
        \includegraphics[width=1\textwidth]{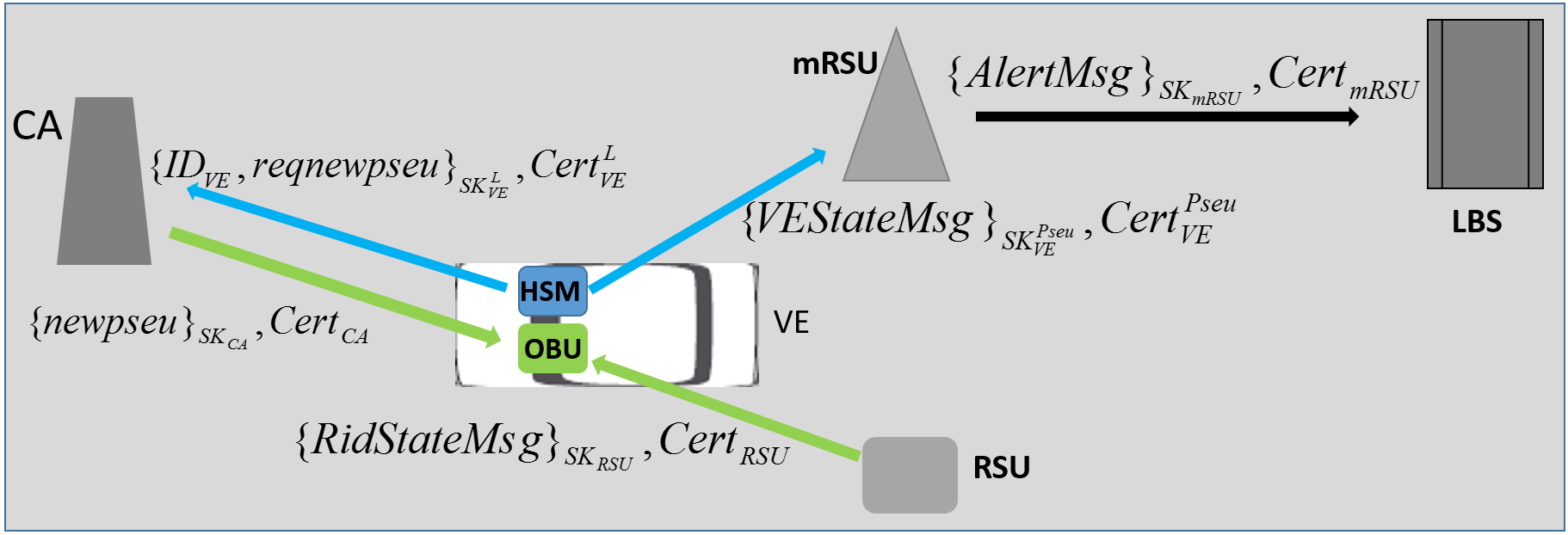}
    \end{center}
    \caption{\textit{The secured wireless communication concept implemented by the system. CA is the corresponding certificate authority for issuing public key certificates, RSU is a secondary roadside unit, mRSU is the main and more powerful roadside unit, while LBS denotes the local base station, reponsible for controlling the traffic lights at a given intersection. LBS can be the corresponding controller stations of SCATS, SCOOT or InSync.}}
    \label{fig:comstruct}
\end{figure}

\begin{figure}[htb!]
    \begin{center}
        \includegraphics[width=1\textwidth]{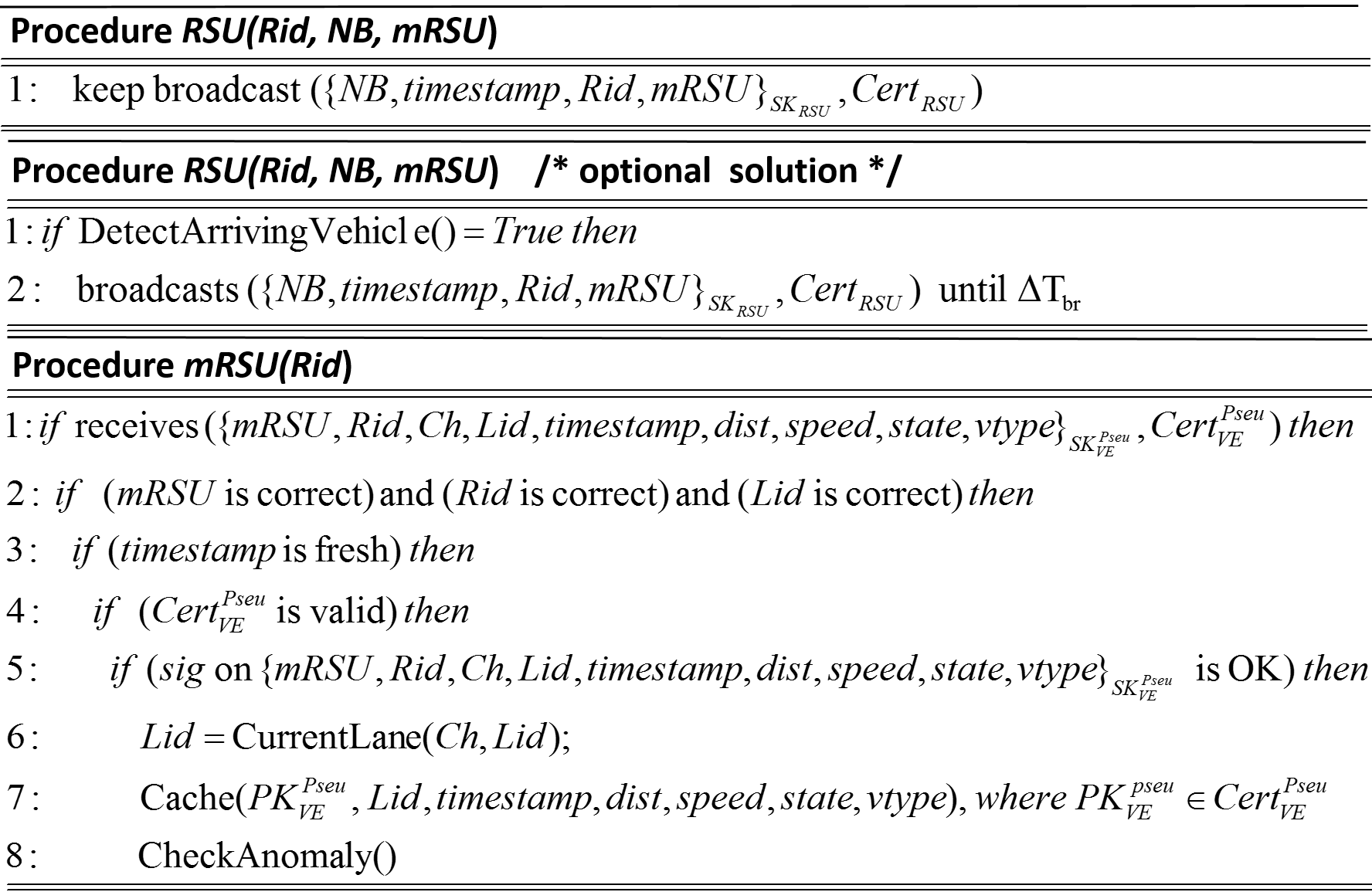}
    \end{center}
    \caption{\textit{The pseudocode of each Roadside Unit, RSU}.}
    \label{fig:RSUpseudo1}
\end{figure}

A RSU periodically broadcasts messages   with the information about the road segment such as road ID (\textit{Rid}), the timestamp, the connections of the neighbor roads and this road at a given intersection (\textit{NB}), and the address of the main RSU (\textit{mRSU}),  all  signed with its private keys with the certificate of the corresponding public key attached. \textit{NB} is a set of 4-tuples, \{(\textit{Rid}$_j$, \textit{Lid}$_j$,$I_{\alpha}$, \textit{Lid}$_i$)\}, as discussed above. We shorthand this message by \{\textit{RidStateMsg}\}$_{SK_{RSU}}$ for brevity (see Figure~\ref{fig:comstruct}). Optionally, each road segment can deploy a sensor device/module at the beginning part to detect the presence of the arriving vehicle, and the corresponding RSU only starts to broadcast  \{\textit{RidStateMsg}\}$_{SK_{RSU}}$ for certain time period $\triangle T_{br}$ when arriving vehicle is detected (see Figure~\ref{fig:RSUpseudo1}). This solution reduces computation overhead for the RSUs. Although, we put it as ``optional" because of two issues: (i) the cost of installing additional vehicle detection module at the beginning of each road segment can impose  additional cost, and (ii) due to the speed of the arriving vehicles can be high, vehicle will get out from the communication range of the RSUs before it gets the \{\textit{RidStateMsg}\}$_{SK_{RSU}}$.     Hence, to make sure that vehicles receive \textit{RidStateMsg}, the vehicle detection modules should be located closer to the start of the road segments than the RSUs responsible for broadcasting. 

\begin{figure}[htb!]
    \begin{center}
        \includegraphics[width=1\textwidth]{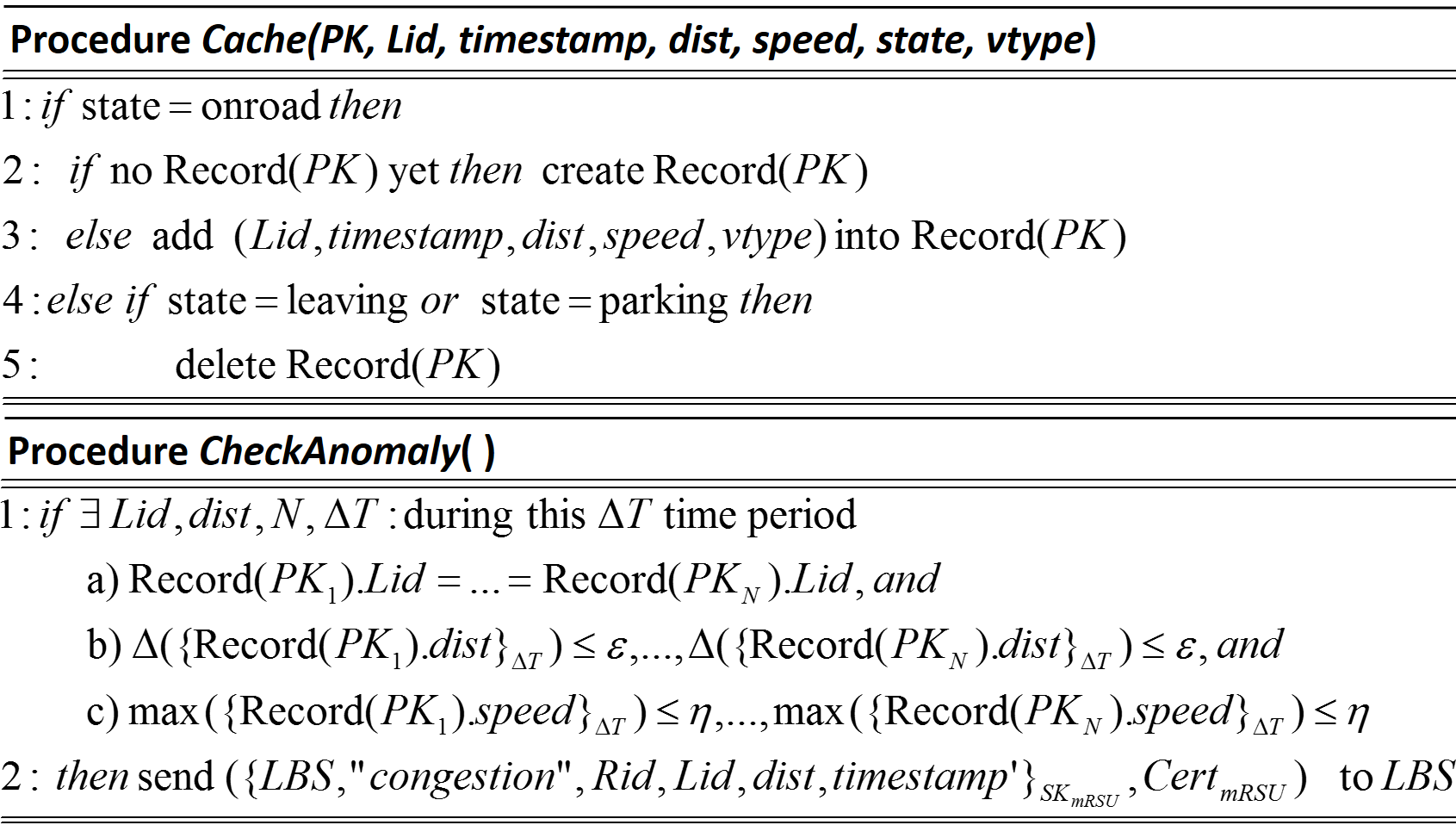}
    \end{center}
    \caption{\textit{The pseudocode of the} \textit{Cache} \textit{and} \textit{CheckAnomaly} \textit{procedures}.}
    \label{fig:RSUpseudo2}
\end{figure}

\begin{figure}[htb!]
    \begin{center}
        \includegraphics[width=1\textwidth]{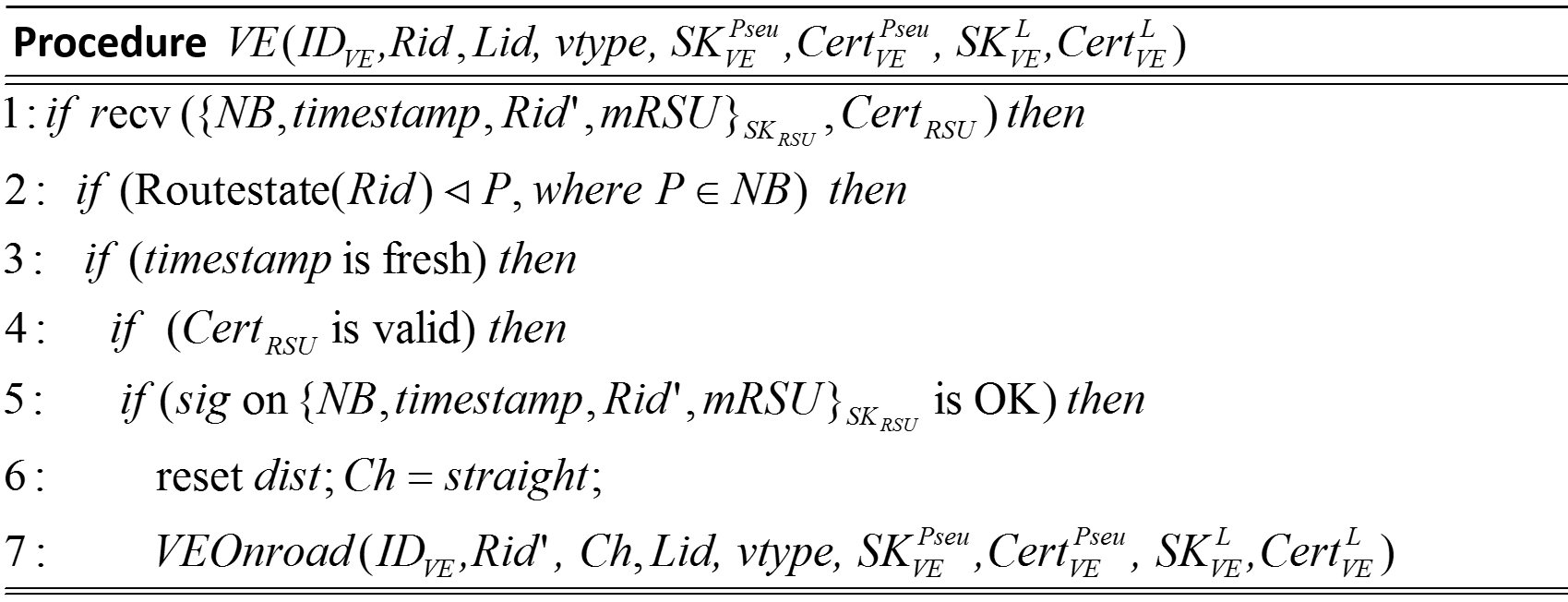}
    \end{center}
    \caption{\textit{The pseudocode of each vehicle, VE}.}
    \label{fig:VEpseudo1}
\end{figure}

When a vehicle \textit{VE} receives \{\textit{RidStateMsg}\}$_{SK_{RSU}}$, it verifies the set of information about the  surrounding roads at a given intersection (see Figure~\ref{fig:VEpseudo1}), and checks if this is consistent with its routestate, namely, it really arrives at the correct road segment. Then, it verifies the freshness of the message along with the validity of the public key of the RSU, based on the certification. This is followed by verifying the signature using the attached public key. In case of success, \textit{VE} sets its state to ``onroad", resets the distance \textit{dist}, and starts sending its status message signed with the pseudo key (the most recent short-term pivate key) to mRSU, namely,  \{\textit{VEStateMsg}\}$_{SK^{Pseu}_{VE}}$, attached the certificate of the corresponding short-term public key (see Figure~\ref{fig:comstruct}). Note that the address/ID of the corresponding mRSU is included in \{\textit{RidStateMsg}\}$_{SK_{RSU}}$, sent by the RSU. \textit{RidStateMsg} contains details of the road segment, a fresh timestamp, the travelled distance, \textit{dist}, so far from the beginning of the road segment, the speed, and the state (see Figure~\ref{fig:RSUpseudo2}).               Each vehicle also has a type (represented by \textit{vtype}), which distinguishes emergency vehicles in action (e.g., police, fire engine, ambulance) from normal vehicles, as well as public transport vehicles. 

In case vehicle \textit{VE} has just parked/parking on the road segment \textit{Rid}, it will set the state to ``parking", then periodically send a message signalling the status parking to mRSU. Upon receiving this message and  successful verifications, mRSU will delete the corresponding record.

mRSUs are responsible for more computations than RSUs, e.g., signature verification, as well as message storage for a certain time period, and anomaly detection based on the stored traffic messages. Once receives (\{\textit{VEStateMsg}\}$_{SK^{Pseu}_{VE}}$, Cert$^{Pseu}_{VE}$), mRSU checks if it is the addressee, the parameters of the road segments are correct, and the timestamp is fresh. Then, it verifies the attached public key certificate and the signature, then caches the message upon successful verification. In case there is no any record created for $PK$ (the corresponding short-term public key of $SK^{Pseu}_{VE}$) yet, mRSU creates a new record and adds the new message to the record. Once the RSU at the end of the road segment signals the leaving of the vehicle with $PK$, mRSU will delete the corresponding record to free the memory. We recall that the vehicles will change to  another pair of short-term keys (pseudonyms) only at the intersections, after leaving the previous road segment, hence, preventing the RSUs and mRSUs from tracking certain vehicles. Vehicles use different unlinkable keys on different road segments. On the other hand, since $PK$ will not be changed during a road segment, all of the messages of the corresponding \textit{VE} will be cached by the mRSU, and hence, the anomaly detection will be precise. Hence, this concept will allow vehicles to be traceable during a single road segment.  
          
The anomaly detection algorithm is based on the cached messages of a certain number ($N$) of vehicles, which get stuck at a certain road segment area (determined by \textit{dist}), within a pre-defined time period $\triangle T$. A congestion is detected if during this $\triangle T$, the number of vehicles within the constrained area (with radius $\epsilon$) of the road segment is $N$, and their maximum speed is below a certain treshold $\eta$. This case an alert will be sent the local base station LBS by mRSU. The intuition is that congestion is characterized by a number of vehicles got stuck at a certain point of a road segment for a while. If either $N$ or $\triangle T$ is small, then the congestion is not critical and no alert will be sent. Note, again, that in our case, LBS can be the regional/local station defined in SCATS, SCOOT or InSync, but now the optimisation/calculation also takes into account the alert messages sent by the mRSUs.     

\begin{figure}[htb!]
    \begin{center}
        \includegraphics[width=1\textwidth]{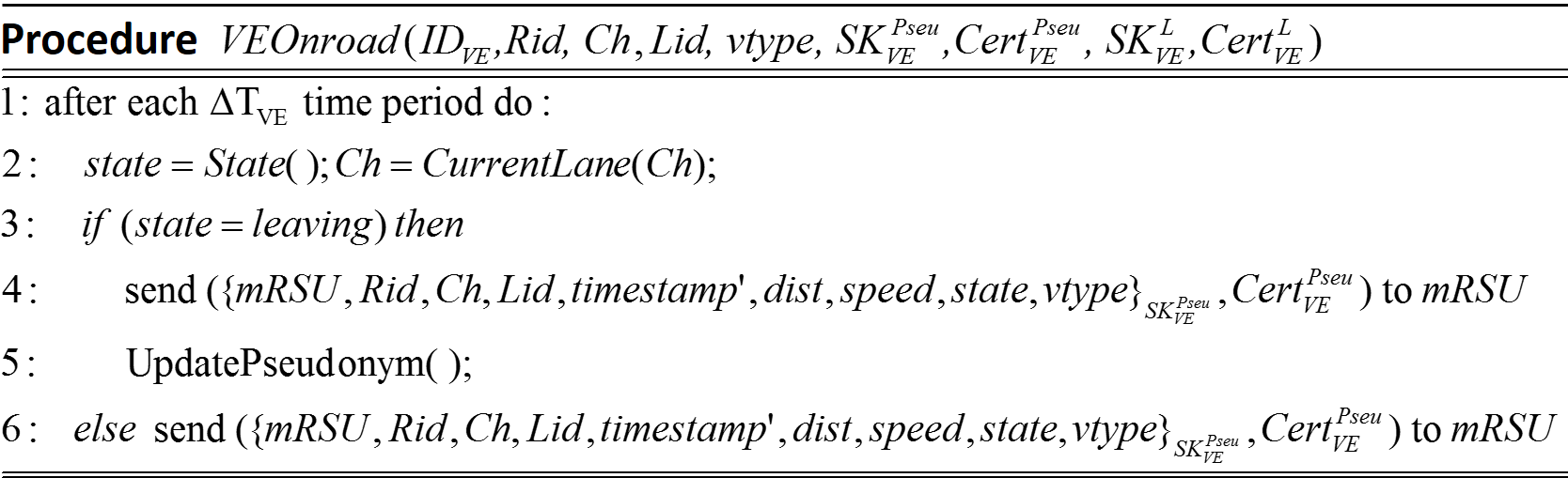}
    \end{center}
    \caption{\textit{The pseudocode of the VEOnroad procedure.}}
    \label{fig:VEpseudo2}
\end{figure}

\begin{figure}[htb!]
    \begin{center}
        \includegraphics[width=0.8\textwidth]{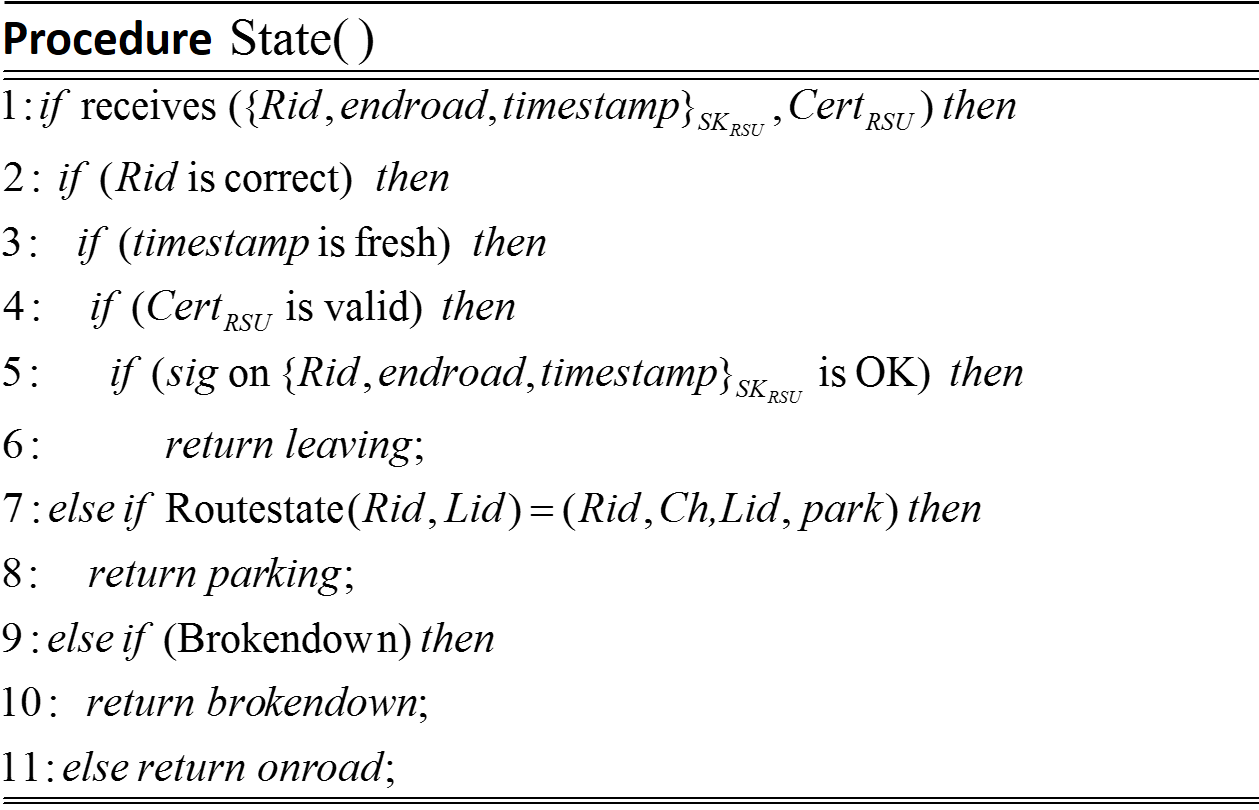}
    \end{center}
    \caption{\textit{The pseudocode of the State() procedure.}}
    \label{fig:VEState}
\end{figure}

\begin{figure}[htb!]
    \begin{center}
        \includegraphics[width=0.8\textwidth]{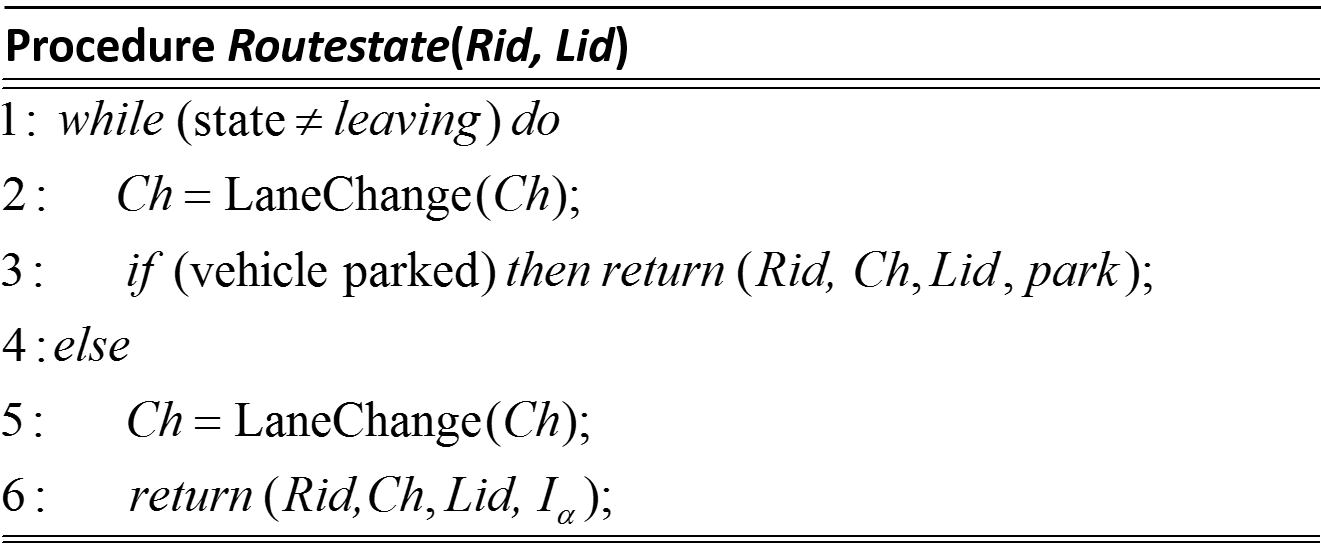}
    \end{center}
    \caption{\textit{The pseudocode of the Routestate procedure.}}
    \label{fig:VEpseudo3}
\end{figure}

\begin{figure}[htb!]
    \begin{center}
        \includegraphics[width=1\textwidth]{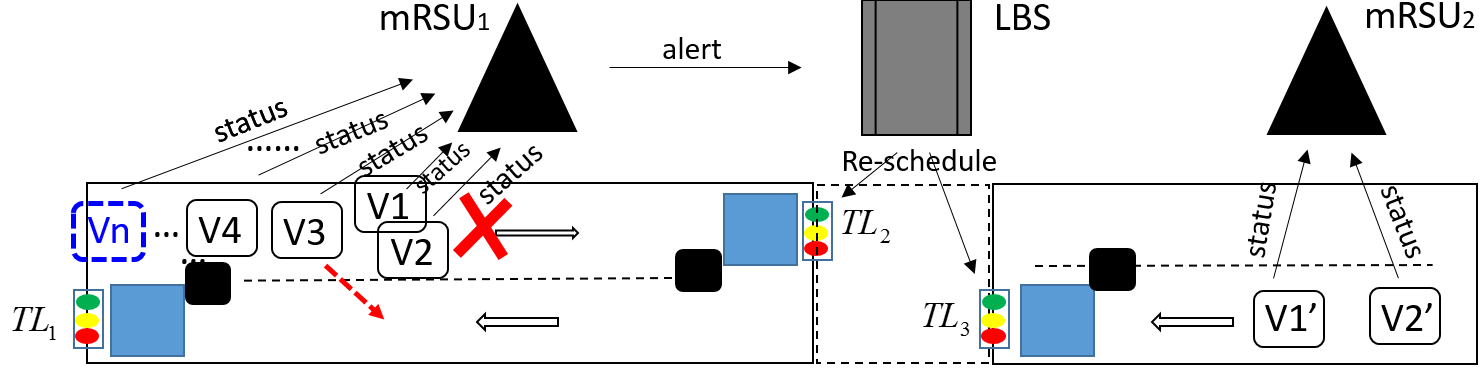}
    \end{center}
    \caption{\textit{The accident scenario in Figure~\ref{fig:accident1},  with the proposed alerting system.}}
    \label{fig:accident2}
\end{figure}

When the local base station LBS dedicated to this intersection receives an alert, it analyses the traffic status in all the road segments in this intersection, and re-schedule the traffic lights based on an optimisation algorithm. LBSs also forward their information about the congestion to neighboring LBSs. There can be several scenarios, for instance, let us consider Figure \ref{fig:accident2}, where we can see the accident example in Figure \ref{fig:accident1}, but now with our proposed solution. In this case, there is only two vehicles on the opposite road segment without any incident alert, hence, LBS will send instruction to decrease the green time of \textit{TL}$_3$ and increase the green time of \textit{TL}$_2$. In addition, LBS forwards this information to the neighbor LBSs as well to extend the optimisation. On the other hand, in the scenario depicted in Figure~\ref{fig:accident2}, if the LSB$_1$ receives that the road opposite with the alerted one also gets stuck, then the information is forwarded to the neighbor LSBs, and in this case, LSB$_2$ will divert all the vehicles (\textit{Vm}, \textit{Vk}, \dots) to the crossing road segment.               

\begin{figure}[htb!]
    \begin{center}
        \includegraphics[width=1\textwidth]{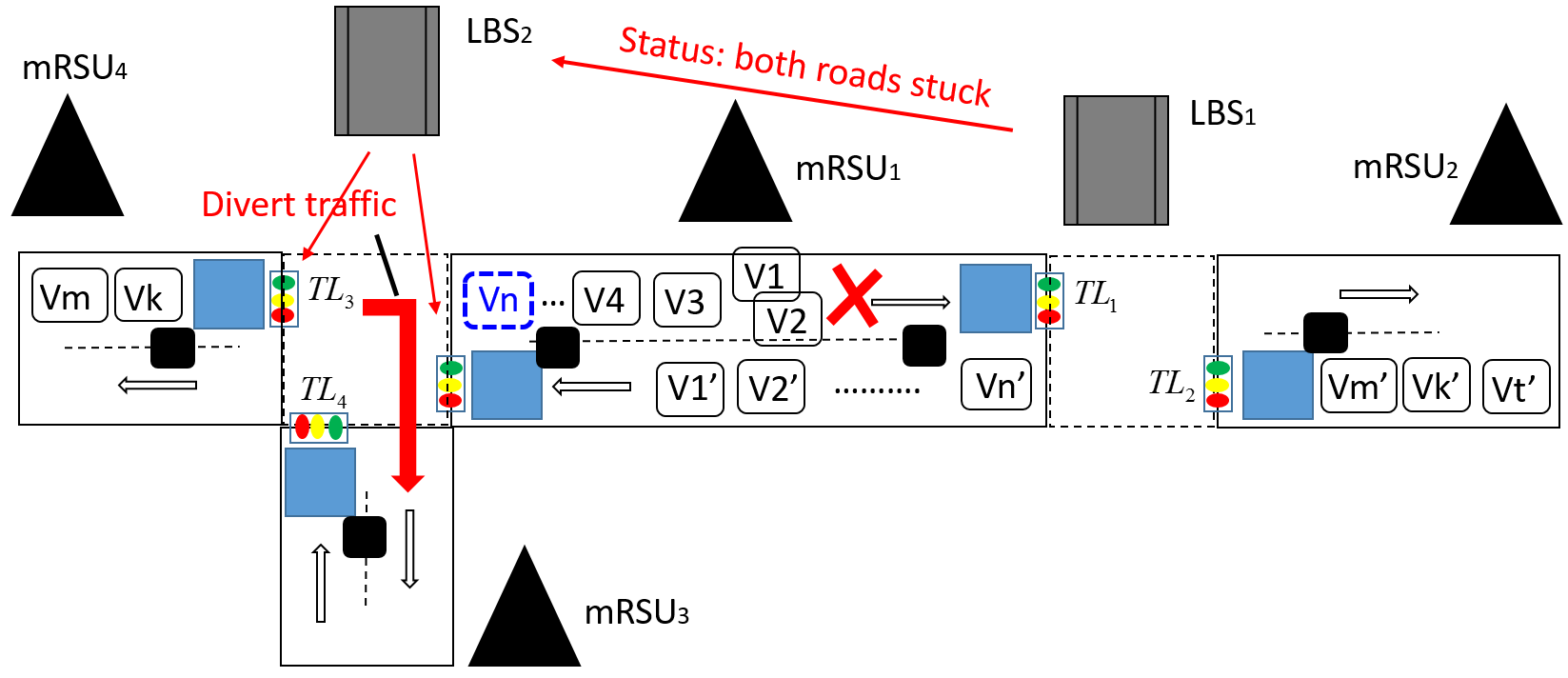}
    \end{center}
    \caption{\textit{A scenario when both the two directions got stuck, and  LBS$_2$ diverts the traffic.}}
    \label{fig:divert1}
\end{figure}

\section{A Secure Traffic Congestion Alert Concept (Type \textit{S2})}
\label{sec:S2}

Similar to the approach of type \textit{S1} in Section~\ref{sec:S1}, the approach \textit{S2} is also a solution  
on top of the existing ATCSs, but here we assume that vehicles are equipped with precise positioning devices. Hence, the vehicles are be able to send their recent position to the RSUs and mRSUs, which also means that no vehicle path and lane tracking modules (Figure~\ref{fig:VEpath}) are requied.  Otherwise, the system architecture and  settings of this approach is similar to the settings in Section~\ref{sec:S1}.    

\begin{figure}[htb!]
    \begin{center}
        \includegraphics[width=1\textwidth]{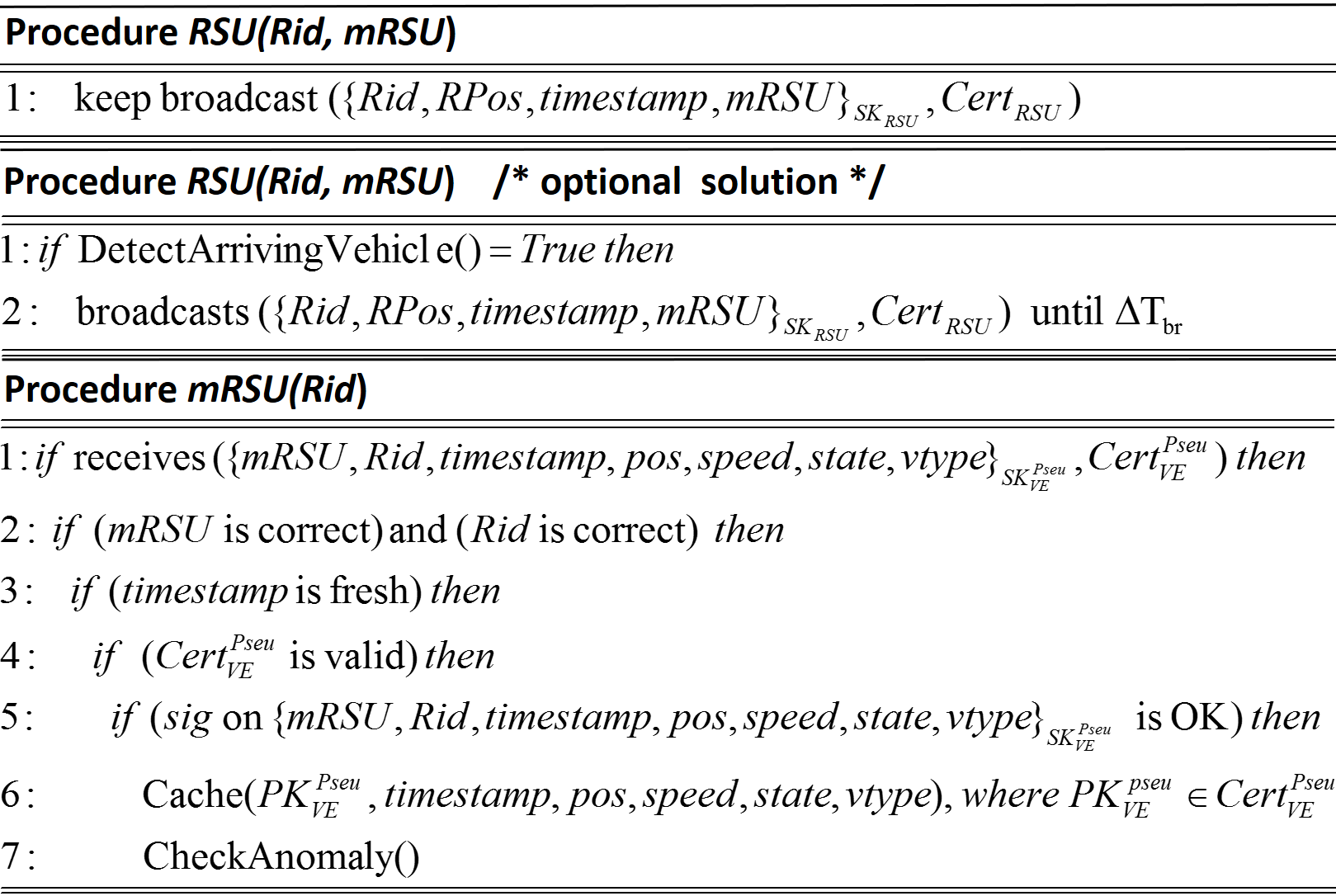}
    \end{center}
    \caption{\textit{The pseudocode of each Roadside Unit with precise positioning devices, RSU}.}
    \label{fig:RSUpseudo1pos}
\end{figure}

In this case, the RSU broadcasts periodically the signed message including the road segment ID, \textit{Rid}, the location of the road (\textit{RPos}), the       
timestamp, and the ID of the corresponding \textit{mRSU}. The \textit{mRSU} now expects a message with the vehicles current position, speed, state and type. Upon successful verification of the IDs, the timestamp and the signature, mRSU will cache the message, and checks for anomaly (i.e., congestion). The congestion detection is now based on the coordinate capturing the road-direction   (for monitoring the vehicles' progress). In case of any congestion is detected, the mRSU will send an alert message to the local base station (LBS). 
In this case, the local base station will    determine the lane on which the congestion happens, based on the received location information, \textit{pos}, and the lane coordinates (\textit{LPos}).     We assume that each LBS stores a record about the \textit{Rid}s and the corresponding \textit{LPos}. 

\begin{figure}[htb!]
    \begin{center}
        \includegraphics[width=1\textwidth]{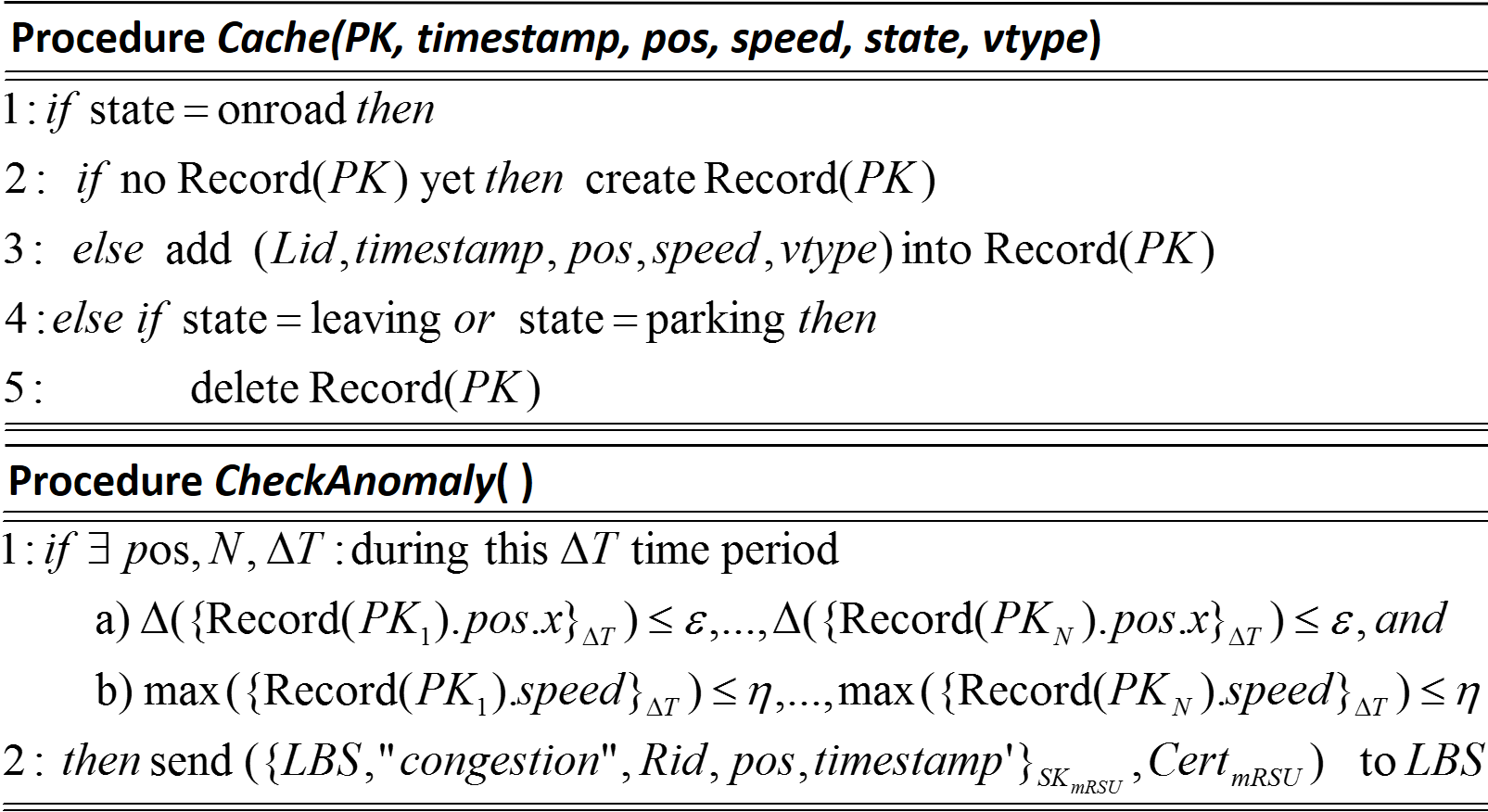}
    \end{center}
    \caption{\textit{The pseudocode of the} \textit{Cache} \textit{and} \textit{CheckAnomaly} \textit{procedures}.}
    \label{fig:RSUpseudo2pos}
\end{figure} 

The vehicles are equipped with precise positioning devices, and are aware of their current position, \textit{pos}. After receiving the message from RSU, the vehicles will check if its current location is in the road (characterized by \textit{RPos}), then the freshness and the signature of the message. Otherwise, the message will be ignored. Upon successful verification, after each $\triangle T_{VE}$ time period,  the vehicles periodically send a signed message with their current speed, location, state, and type to the corresponding mRSUs.       
          
\begin{figure}[htb!]
    \begin{center}
        \includegraphics[width=0.8\textwidth]{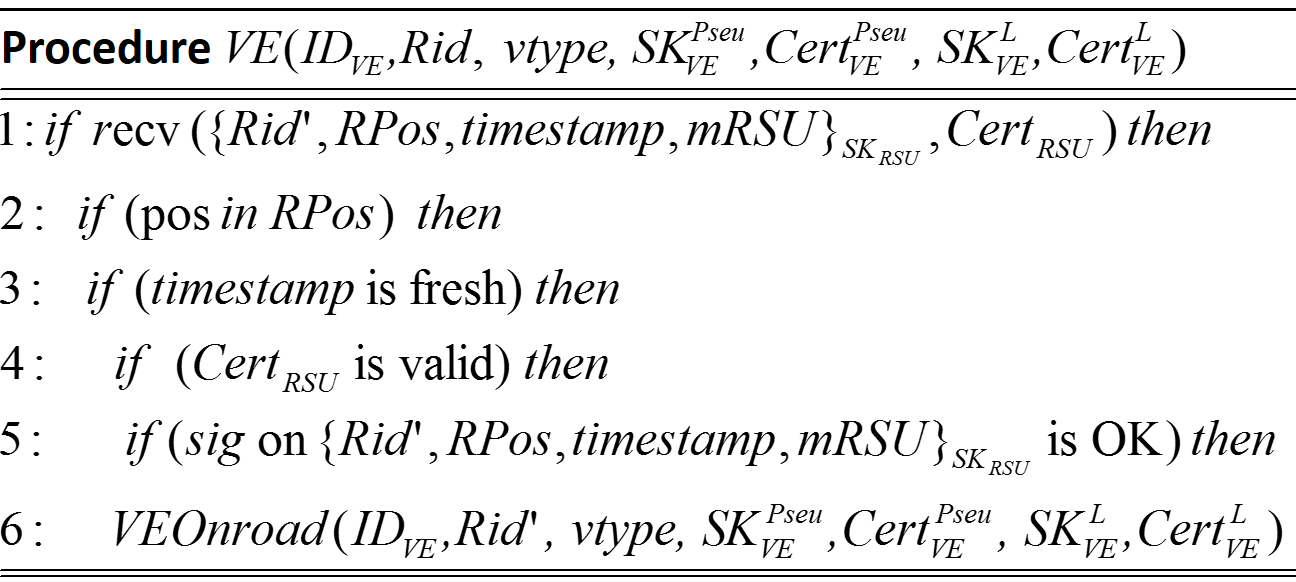}
    \end{center}
    \caption{\textit{The pseudocode of each vehicle, VE, with positioning devices}.}
    \label{fig:VEpseudo1pos}
\end{figure}    

\begin{figure}[htb!]
    \begin{center}
        \includegraphics[width=1\textwidth]{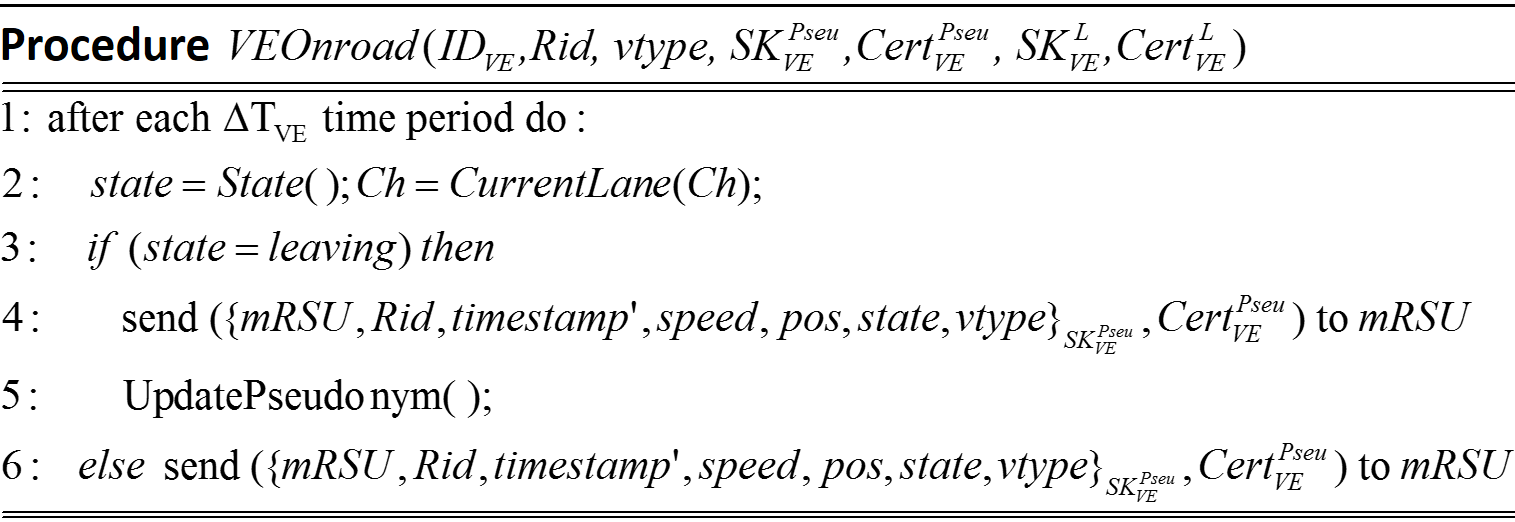}
    \end{center}
    \caption{\textit{The pseudocode of the VEOnroad procedure with positioning devices.}}
    \label{fig:VEpseudo2pos}
\end{figure}

\begin{figure}[htb!]
    \begin{center}
        \includegraphics[width=0.8\textwidth]{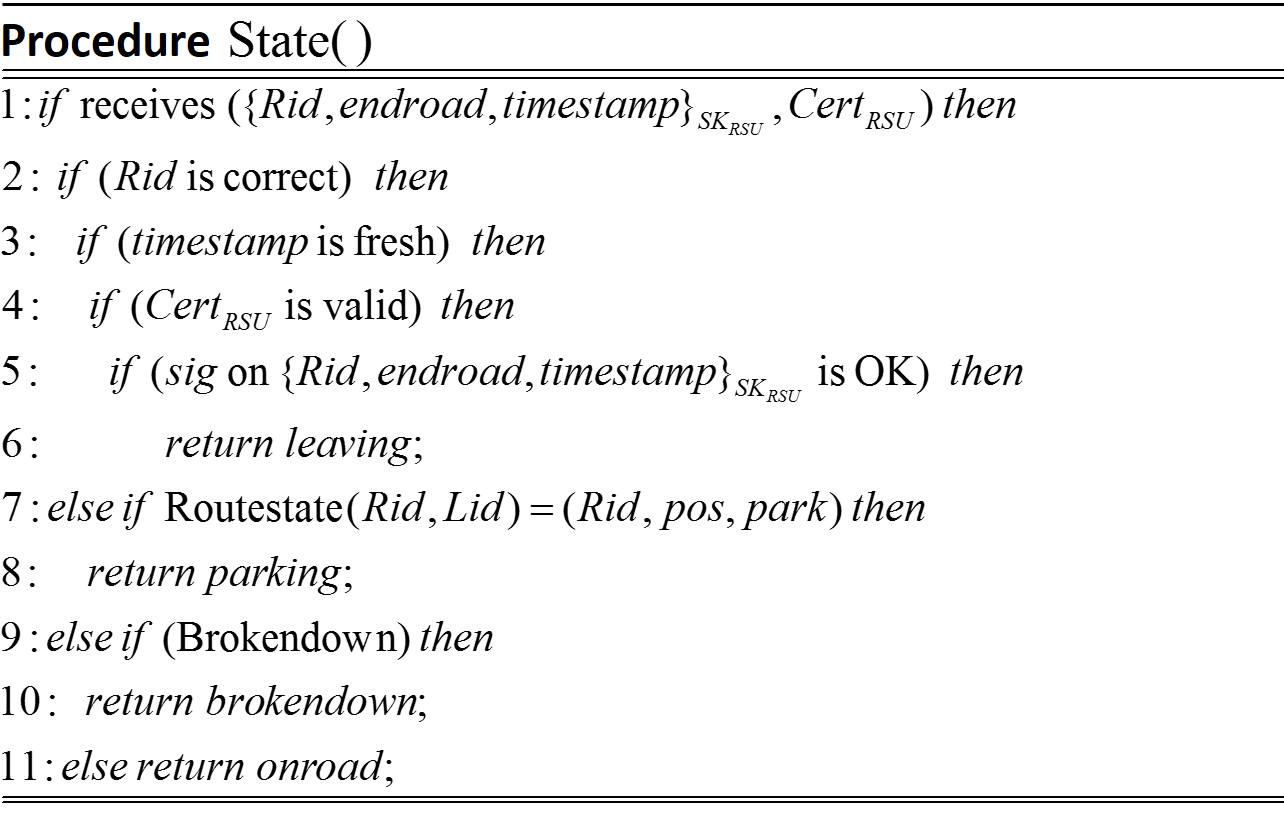}
    \end{center}
    \caption{\textit{The pseudocode of the State() procedure with positioning devices.}}
    \label{fig:VEStatePos}
\end{figure}

\begin{figure}[htb!]
    \begin{center}
        \includegraphics[width=0.7\textwidth]{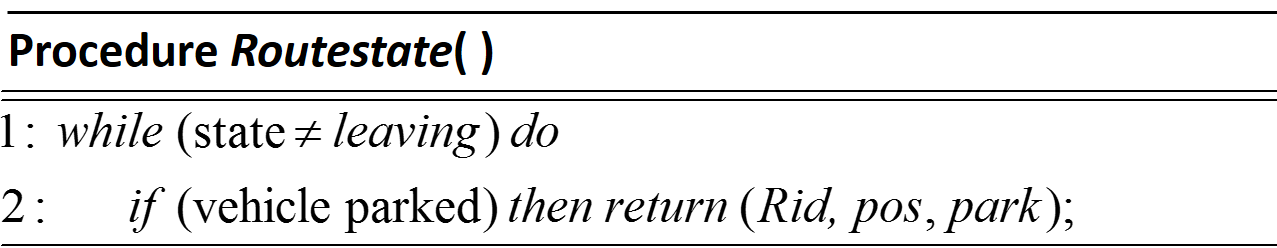}
    \end{center}
    \caption{\textit{The pseudocode of the Routestate procedure with positioning devices.}}
    \label{fig:VEpseudo3pos}
\end{figure}

The routestate corresponding to a vehicle is defined by the triple, and specifies the current state of the vehicles. Here, we only define routestate to distinguish parking and active vehicles. The corresponding algorithms will be changed accordingly as follows:     

\section{Security Analysis and Discussion}
In this section, we discuss the security and privacy properties of our proposed solutions, taking into account the points given in Section~\ref{sec:adversary}. In the first part, we focus on the possible attacks targetting the wireless communication protocol.  We conclude that attacks can only be successful by exploiting potential  weaknesses found in the settings of HSMs, OBUs, RSUs, mRSUs and LBSs (e.g., operating system, software, firmware, or physical protection). In the second part, we examine the case of malware-based attacks, via   vehicular botnets.

We distinguish two classes of attackers, internal and external attackers.
We modelled and verified the communication protocol and the API of the HSMs using the ProVerif protocol verification tool \cite{proverif}.  

The messages are digitally signed using the private keys, hence, in order to get the fake location and speed information cached at the mRSUs the attackers have to successfully forge the digital signatures. However this is hard as we assumed that the HSMs responsible for private keys/digital signature generation is tamper resistant, and, we proved, using the ProVerif tool, that their APIs (Application Programming Interface) are secure (i.e., they never give out the private keys in plaintext).     

The timestamps provided by the hardware security modules (HSMs) are included in the messages to prevent  message replay attacks. Although using timestamp requires precise time source and synchronisation, it is still more convenient than random nonce (number used once) in that it does not require challenge-response communication, which can be critical in case of vehicle with high speed, as the communication may not  finished before the vehicles leave the communication range.  Hence, in order to replay old messages attackers should either generate valid signatures on the message of with fresh timestamp or change the time source in the vehicle, which is hard as we assumed that the HSMs responsible for time source and private keys/digital signature generation is tamper resistant, and again, we proved using the ProVerif tool that their APIs  never give out the private keys in plaintext.  

From privacy perspective, we assume that the vehicles regularly changing the public and private keypairs at the intersections (before arriving to a new road segment), and no future keypairs can be linked to the old ones. This will make difficult for an external observer who eavesdrop the wireless communication to track the vehicles for a long-term. It also makes more difficult for the system administrator/technician who has access to the mRSUs to track the vehicles. Of course, our congestion alert concept is not strictly linked with the pseudonyms. We only choose to apply pseudonyms because it has been widely proposed. Our concept can be easily incorporated with different privacy enhancing approaches, e.g., group signature \cite{group}, as well. 

Denial of Services (DoS) attacks can be launched against our system, when the attackers send a huge amount of messages to the mRSUs. These messages are not valid messages but consume the resource of the mRSUs, as they have to check the signatures and then drop the messages. Different methods against DoS attacks~\cite{denial1, denial2, denial3, denial4} can be implemented at the mRSUs to mitigate their impact.    

Finally, jamming attack is always a potential threat in wireless communication, where the attacker prevent the RSUs and mRSUs to receive the messages sent by the vehicles by causing interference with powerful antennas. Different anti-jamming techniques~\cite{jamming1} can be adopted to mitigate the effect of jamming. This could be part of future research directions. 

\textit{\textbf{Vehicular Botnet}}: Let assume that either the vehicles' sensors or HSMs, OBUs can be infected 
with some kind of malwares, and hence, the attackers can achieve that their controlled vehicles will send correcly signed messages on incorrect information. The attackers' ultimate goal now is to buld up an extensive botnet of vehicles, making them sending information indicating a fake traffic congestion (or several congestions on different roads).      For instance, the attacker who control a botnet of vehicles can change the location and speed data of the infected vehicles, and then calling the HSM API to sign and send messages containing this fake information. By reporting several fake traffic congestions to the LBSs, the attackers will be able to fool the system, potentially leading to real congestions. 

An interesting research direction could be investigating the mitigation methods for this type of attack, perhaps applying machine learning or examining the problem from game-theoretic aspects.  

\section{Conclusion and Future Works}
In this position paper, we discussed the most relevant and broadly used adaptive traffic control systems, including 
SCATS, SCOOT and InSync. We argue that future vehicular communication technologies and infrastructures can be incorporated to these system to provide efficient automated road congestion alarm systems, facilitating more efficient traffic signal control. We highlighted and discussed the security of the existing systems and the possible vulnerabilities and attacks. We also discussed some open research directions and possible approaches. Finally, we provide theorectical architecture and communication protocols on top of the existing systems (SCATS, SCOOT, InSync) for two possible automated road congestion alarm systems incorporating vehicle-to-infrastructure communication technologies.   

As a potential future work, we will develop a Java based simulation framework, specifically designed to simulate and compare the effectiveness of the proposed extension with the basic SCATs, SCOOT and InSync solutions. As a longer term plan a testbed will be deployed to analyse the effectiveness of the proposed methods. 
 
\bibliographystyle{abbrv}
\bibliography{traffic}

\end{document}